\documentclass[%
 preprint,
 amsmath,amssymb,
 aps,
]{revtex4-2}

\usepackage{graphicx}
\usepackage{bm}
\usepackage{moreverb}
\usepackage{graphicx}
\usepackage{dcolumn}
\usepackage{bm}
\usepackage[version=3]{mhchem} 
\usepackage{color, soul}
\usepackage{multirow}
\usepackage{comment}
\usepackage{booktabs}
\usepackage{makecell}
\usepackage{hyperref}

\begin{document}
\preprint{APS}

\title{Structural Influence on Exciton Formation and the Critical Role of Dark Excitons in Polymeric Carbon Nitrides}

\author{Changbin Im}
\affiliation{Institute of Electrochemistry, Ulm University, Albert-Einstein-Allee 47, 89081 Ulm, Germany.}

\author{Radim Beranek}
\affiliation{Institute of Electrochemistry, Ulm University, Albert-Einstein-Allee 47, 89081 Ulm, Germany.}%

\author{Timo Jacob}%
\affiliation{Institute of Electrochemistry, Ulm University, Albert-Einstein-Allee 47, 89081 Ulm, Germany.}%
\affiliation{Helmholtz-Institute-Ulm, Helmholtzstr. 11, 89081, Ulm, Germany}
\affiliation{Karlsruhe Institute of Technology (KIT), P.O. Box 3640, 76021, Karlsruhe, Germany}
\date{\today}

\begin{abstract}
Polymeric carbon nitrides (PCNs)
exhibit intriguing optical properties and exceptional performance in (photo)catalysis, optoelectronics, and energy storage. Nevertheless, the intricate
phenomena involving light absorption, formation of long-lived excitons,
photo-charging, and photochemical processes observed in PCNs remain poorly understood.
Our theoretical investigation elucidates the origin of distinct dark and bright excitons, their stability and lifetimes, and their correlation with the microstructural attributes of PCNs.
Furthermore, we identify the decisive role of dark excitons in catalytic reactivity, which underlies the differences in the photocatalytic performance of different PCN derivatives.
\end{abstract}
\maketitle
\textit{Introduction}.---Polymeric carbon nitrides (PCNs), commonly referred to as graphitic carbon
nitrides or melon, along with their (semi)crystalline derivatives such as poly(heptazine imide)
(PHI) have emerged as compelling materials for
photocatalysis, optoelectronics, and energy storage applications.\cite{kessler2017functional,podjaski2021optoelectronics,gouder2023integrated,liu2015,cao2015polymeric} This can be attributed to their straightforward
synthesis, robust chemical stability, and optical properties that span from the UV
to visible light range.\cite{cao2015polymeric} Research on PCNs has demonstrated their efficacy
in photocatalytic applications such as hydrogen production under near-visible light, water
splitting, \ce{CO2} reduction, and \ce{H2O2} production.\cite{cheng2021carbon}
Highly crystalline forms of carbon nitrides are typically obtained using molten salt methods,
resulting in highly ordered 2D structures.\cite{chen2017easier} Particularly, these derivatives, containing cations, are of significant interest due to enhanced photocatalytic activity and the photo-charging behavior that can be utilized in time-delayed photocatalysis or boosted photocurrent response.\cite{lau2017dark,adler2021photodoping,kasap2016solar}
Despite the plethora of useful and intriguing properties, the knowledge of the relationship between the structural and optical properties in PCN materials is still largely underdeveloped.\cite{cao2015polymeric}
Merschjann \textit{et al}. reported that charge transport in PCN structures occurs between interplanar layers.\cite{merschjann2015complementing}
Godin \textit{et al}. discussed the charge trapping phenomena in conventional PCNs, and Yang \textit{et al}. reported that the excess electron accumulation in ionic PCNs leads to lower quantum yields due to accelerated electron--hole ($e$--$h$) recombination.\cite{godin2017time,yang2019electron} 
However, any clear conclusions from experimental studies of exciton dynamics in PCNs using time-resolved spectroscopy remain limited due to the lack of detailed understanding of the structure-induced excited state properties.\cite{godin2017time,yang2019electron,li2021ultrafast}
For example, the frequently observed trapped excitons in PCNs are typically attributed to structural defects.
Nonetheless, there is a notable absence of detailed discussion on the nature of these structural defects or their impact on photocatalytic activity.
Utilizing many-body perturbation theory based on Green's function formalism can provide accurate descriptions of excited state properties such as accurate bandgap, exciton energy, and absorption spectra. Our previous studies reported strong exciton binding energy,\cite{Wei2013} and have examined the effects of PCN microstructures, such as corrugation, degree of condensation, and stacking, focusing on their thermodynamic stability, electronic structure, and optical properties.\cite{im2023structure} We particularly noted that the interlayer interactions in PCNs play a significant role in both thermodynamic and electronic properties.\cite{im2024unraveling}
Herein, we perform first-principles calculations based on \textit{GW} and the Bethe--Salpeter equation (\textit{GW}-BSE),\cite{rohlfing1998electron} to elucidate the crucial link between the various structural motifs in PCNs and the corresponding optical properties, formation of bright and dark excitons, exciton stability and lifetime, as the resulting factors governing the photocatalytic activity. We demonstrate that the strongly localized electrons in PCNs induce active interlayer transitions due to the stacked structure of PCNs. Furthermore, we show that the interactions between heptazine units vary with the microstructure of PCNs, affecting the formation and the lifetimes of the dark excitons more significantly than of the bright excitons. Our findings thus establish the pronounced structure-related variations of the dark excitons as a key factor determining the experimentally encountered optical properties and photocatalytic performance of different types of PCNs.

\textit{Method}.---All \textit{ab initio} calculations were performed using VASP (6.4.1),
which uses the projector augmented wave potentials (PAWs).\cite{hafner2008ab} We considered
PCN structures optimized by the local density approximation (LDA) and generalized
gradient approximation (GGA) functionals, respectively, to compare the description of the highly-localized
$\pi$-electrons within the heptazine units.\cite{im2024unraveling} The Perdew--Zunger 
parametrization of Ceperley--Alder Monte-Carlo correlation (CA) functional\cite{pzca1,pzca2} with the
many-body dispersion energy method (MBD@rsSCS)\cite{mbd1,mbd2} is used within LDA, while PBE-D3 is adopted
for the GGA calculations.\cite{perdew1996generalized,grimme2010consistent} To describe the
single-particle excitations of the PCN structure, we conduct a single-shot \textit{GW}
calculation to avoid bandgap overestimation.\cite{hedin1965new,gw2006}
In brief, the \textit{GW} approximation neglects the vertex functions in Hedin's formalisms,
and the self-energy ($\Sigma$) is given as the product of the Green function ($G$) with the
screened self interactions ($W$).
Then, the quasi-particle energy is given by:
\begin{equation}
E^{\mathrm{QP}}_{n\textbf{k}} = E_{n\textbf{k}} + Z_{n\textbf{k}} \langle \psi_{n\textbf{k}} |
\Sigma - V_{\mathrm{xc}} | \psi_{n\mathrm{\textbf{k}}} \rangle,
\end{equation} \label{equ1} 
where $\textbf{k}$ and $n$ are the wave vector and band index, $Z_{n\textbf{k}}$ is the
renormalization function obtained from the self-energy with DFT eigenvalues.\cite{ljungberg2015cubic}
In the $G_{0}W_{0}$ calculations, we include 256 bands for the 2D PCN-structure and 512 bands
for 3D PCN-structure, respectively. For the 2D and 3D PCN structures, $\Gamma$-centered
$2 \times 2 \times 1$ and $2 \times 2 \times 2$ $\textbf{k}$-space meshes are used, respectively. 
The quasi-particle energy is updated by using BSE to include \textit{e}--\textit{h} interactions
within the Tamm--Dancoff approximation (TDA)\cite{sander2015beyond}, given by:
\begin{equation}
(E^{\mathrm{QP}}_{c\textbf{k}} - E^{\mathrm{QP}}_{v\textbf{k}})A_{vc\textbf{k}} 
+ \sum_{v'c'\textbf{k}'} \langle vc\textbf{k} | K_{e-h} | v'c'\textbf{k}' 
\rangle A_{v'c'\textbf{k}'} = \Omega A_{vc\textbf{k}},
\end{equation} \label{equ2}
where $c$ and $v$ are indices of conduction and valance bands, respectively, $A_{vc\textbf{k}}$
are $e$--$h$ coupling coefficients, $\Omega$ are the BSE eigenvalues, and
$K_{e-h}$ is the $e$--$h$ interaction kernel.\cite{leng2016gw,rohlfing2000electron}
For the BSE calculations we considered 32 eigenvalues. Then, the independent quasi-particle (IP)
energy considering electron--electron (\textit{e}--\textit{e}) interactions is obtained as \textit{GW}@IP, and the
BSE is used to include \textit{e}--\textit{h} interactions, finally denoted as \textit{GW}@BSE. 
The real and the imaginary parts of the macroscopic dielectric functions are computed including
local field effect.\cite{albrecht1998ab} From the calculated oscillator strengths,
the bright excitons are selected having the 5$\%$ of the highest intensity, and the rest are
classified as the dark states. We modified the VASP source code to obtain the transitional dipole
momentum kernel ($\mu_\mathrm{S}$) to calculate the radiative decay rate of the exciton state
$\mathrm{S}$ ($\gamma_\mathrm{S}$) at wavevector $\textbf{Q} = 0$ using the equation,
given by:\cite{palummo2015exciton}
\begin{equation}
\gamma_\mathrm{S}(0) = \tau_\mathrm{S}(0)^{-1} = \frac{8 \pi e^{2} E_{\mathrm{S}}(0)}{\hbar^{2} c }
\frac{\mu^{2}_\mathrm{S}}{A},
\end{equation}
where $E_\mathrm{S}$ is the exciton energy at $\textbf{Q}$, $A$ is the area of the unit cells,
$c$ is the speed of light, and $\tau_\mathrm{S}$ is the exciton lifetime. Furthermore, we calculate
the effective radiative lifetime $\langle \tau_\mathrm{eff}\rangle$ at temperature ($T$) assuming
a neglectable exciton momentum in PCN structures at high symmetry points.\cite{perebeinos2005radiative} 
Therefore, the effective radiative lifetime using the Boltzmann average of $\tau_\mathrm{S}$ is given by:
\begin{equation}
\langle \tau_\mathrm{eff} \rangle^{-1} = \frac{\sum_\mathrm{S} \tau_\mathrm{S}^{-1}
e^{{-E_\mathrm{S}(0)} \mathbin{/} {k_\mathrm{B}T}} }{\sum_\mathrm{S} e^{{-E_\mathrm{S}(0)}
\mathbin{/} {k_\mathrm{B}T}}}
\end{equation}

\begin{figure}[htp]
\centering
\includegraphics[width=0.80\textwidth]{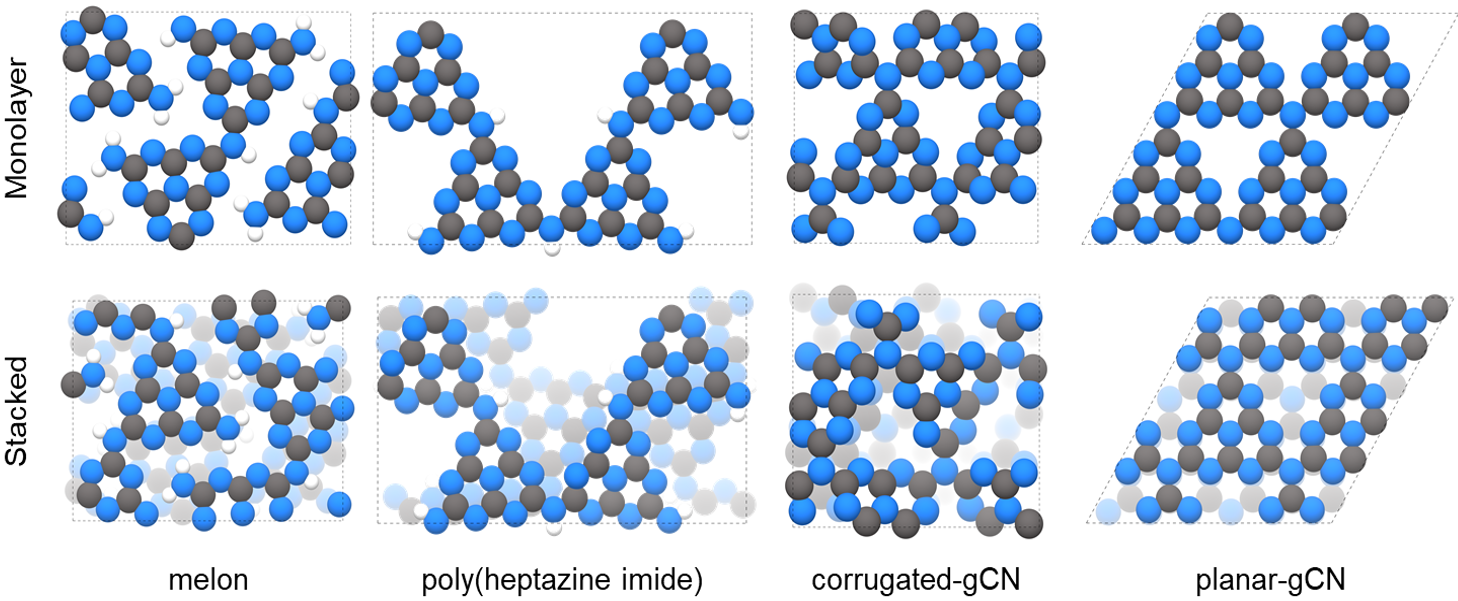}
    \caption{Structures of polymeric carbon nitrides.}
    \label{fig1}
\end{figure}
\textit{Models}.---Figure\,\ref{fig1} illustrates the different PCN structures considered in this work.
These models encompass variations in the degree of poly-condensation of heptazine units, corrugations,
and stacking configurations.\cite{im2023structure} Increased poly-condensation leads to higher carbon
and nitrogen contents, but a reduced hydrogen ratio. The planar-gCN (p-gCN) is known to be energetically
less stable compared to corrugated-gCN (c-gCN), which represents a local minimum in the energy
landscape.\cite{gracia2009corrugated} PCN structures, composed of these heptazine units, typically
exhibit highly-localized $\pi$ electrons. Thus, we primarily present results for optimized geometries
using CA (LDA) functionals. These functionals provide an efficient description of structures with
highly-localized electrons, considered as key properties. To address the limitations of the MBD@rsSCS
dispersion correction, which depends on the exchange--correlation energy from the LDA functionals, 
we also included results obtained with the PBE-D3 \textit{xc}-functional. 
This approach offers a better description of interlayer properties. Results of the optimized geometry
with PBE-D3 are available in the supplementary information (SI) for comparison (Figs.\,S9--S20). 
The lattice parameters of the optimized geometries obtained with both functionals are summarized in
Table\,S1. 

\begin{figure}[htp]
\centering
\includegraphics[width=0.80\textwidth]{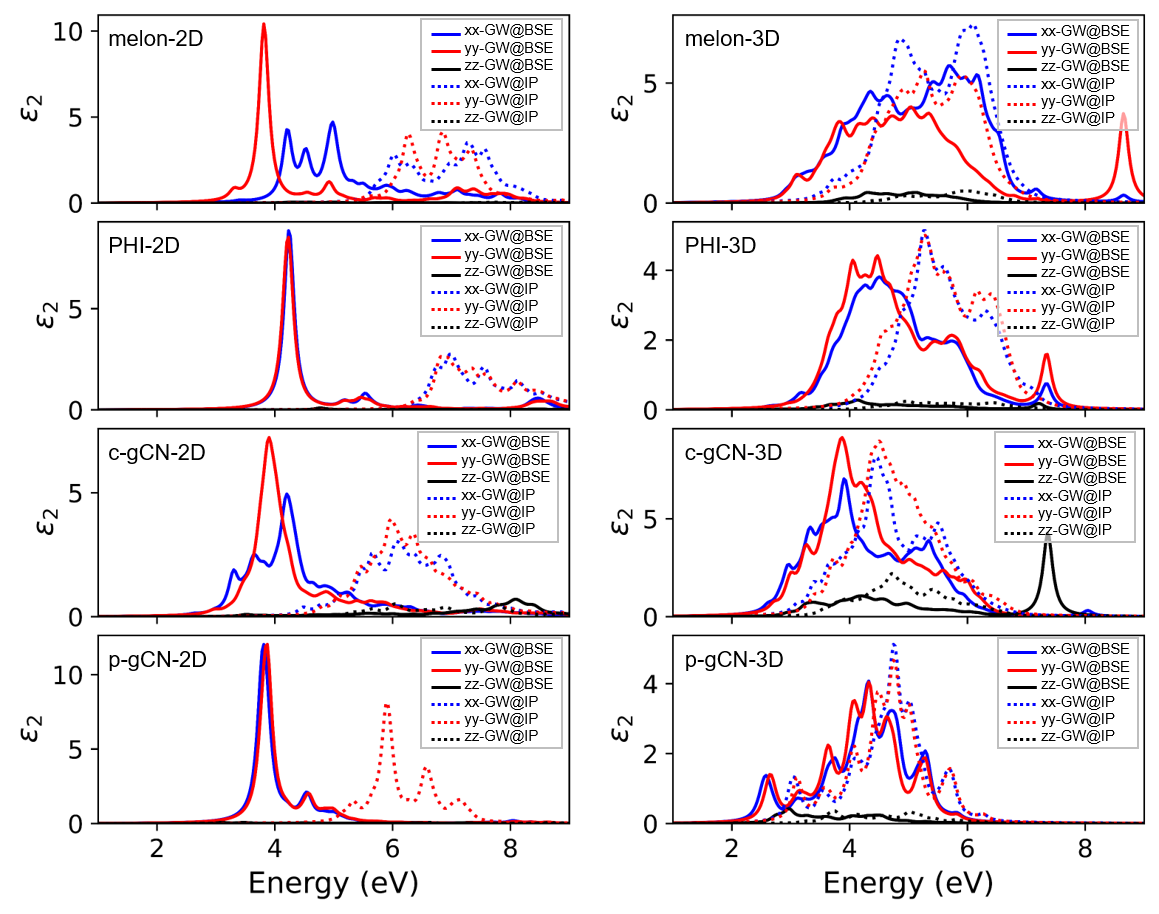}
    \caption{Imaginary part of dielectric functions. 
             The solid and the dashed lines are obtained by the \textit{GW}@BSE and the
             \textit{GW}@IP respectively.}
    \label{fig2}
\end{figure}
\textit{Optical spectra}.---The sole use of density functional theory (DFT) to accurately describe
optical properties has numerous limitations.\cite{hedin1965new,albrecht1998ab}
To address these challenges, we use the BSE to address the energy originating
from \textit{e}--\textit{h} interactions on top of the quasi-particle energy derived from
$GW$ calculations. This $GW$-based quasi-particle energy incorporates precisely screened \textit{e}--\textit{e} interactions, thereby enhancing the accuracy of single excitation property
predictions. Notably, the quasi-particle energy adjusted by the BSE correction typically exhibits a red
shift compared to those predicted by the IP approximation, which omits \textit{e}--\textit{h} interactions.
By analyzing the imaginary part of the macroscopic dielectric functions from both \textit{GW}@BSE and
\textit{GW}@IP calculations, we can estimate the exciton binding energy in 2D and 3D PCN structures,
as illustrated in Figure\,\ref{fig2}. Spatially-projected absorption coefficients reveal distinct patterns:
2D PCN structures show strong exciton binding energy (\textgreater{} 2\,eV).\cite{Wei2013}
However, this significant exciton binding energy diminishes to less than 1\,eV in 3D stacked structures. 
Moreover, the transition from a 2D monolayer to a stacked 3D configuration alters the absorption edges
and reduces the optical anisotropy observed in PHI-2D and c-gCN-2D structures. Since the optical anisotropy
is attributed to the in-plane joint patterns of the heptazine units formed during poly-condensation,
the reduced optical anisotropy implies that the fractions of the 2D PCN micro-structure can indeed interact
with each other throughout the 3D assembly.
In brief, this means that diverse forms of PCN fragments within the PCN structure interact electronically and optically through their stacking configurations, regardless of size or shape.
Our findings provide clear evidence of strong
interlayer interactions, resulting in dramatically decreased exciton binding energy and changes in absorption patterns
including altered optical anisotropy and absorption edges.

\begin{figure}[htp]
\centering
\includegraphics[width=0.80\textwidth]{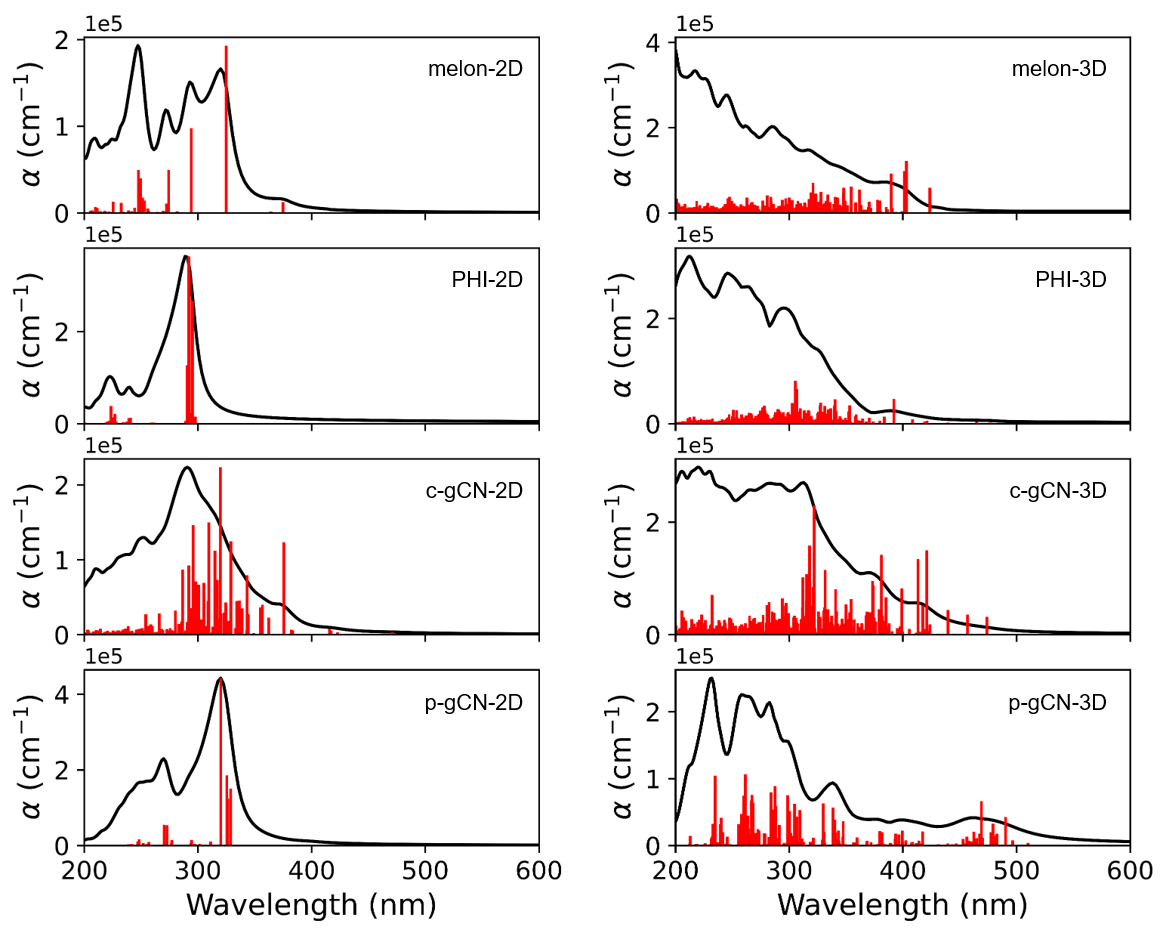}
    \caption{Illustration of the calculated absorption spectra for 2D and 3D PCN models along averaged
    spatial directions. The oscillator strengths in relation to wavelength are presented in red.}
    \label{fig3}
\end{figure}
To better elucidate the origins of the absorption features near the absorption edge, in Figure\,\ref{fig3} we depict the oscillator strengths and averaged absorption spectra along spatial directions for PCN 2D and 3D structures obtained from \textit{GW}@BSE calculations.
Firstly, we observed that each 2D structure of PCNs exhibits distinct absorption spectra patterns. 
When these structures are stacked to form a 3D structure, additional peaks arise due to interlayer interactions.
These emergent peaks are more pronounced in flat structures (melon and p-gCN), where interlayer interactions result in more absorption patterns.
In contrast, in buckled structures (PHI and c-gCN), small absorption peaks appear due to interlayer interactions with the red-shifted 2D peaks.
The strong intensity of absorption peaks near 200\,nm in 3D PCNs is due to the stacking of infinite 2D models, indicating that finite-sized ribbon structures exhibit a reasonable decrease in the intensity of these absorption peaks\cite{im2024unraveling}
The peaks appearing beyond 400\,nm originate from the interactions between layers and the in-plane $\pi$--$\pi^*$ transitions, which will be discussed below.
The absorption edges of the different PCN structures are obtained as 2.93\,eV for melon-3D, 2.60\,eV for PHI-3D, 2.62\,eV for c-gCN-3D, and 1.69\,eV for p-gCN-3D. These PCN structures and bandgaps are completely in line with the experimentally observed bandgaps of PCNs, which range from 2.6 to 2.8\,eV.\cite{cao2015polymeric,butchosa2014carbon,zhang2017optimizing,wang2019increasing}
Especially, the frequently observed long absorption tails can not only be attributed to thermal vibrational motion, but can have a distinct structural origin, primarily from the graphitic (\textit{i.e.} fully condensed) domains.\cite{wang2012polymeric}

\begin{figure}[htp]
\centering
\includegraphics[width=0.9\textwidth]{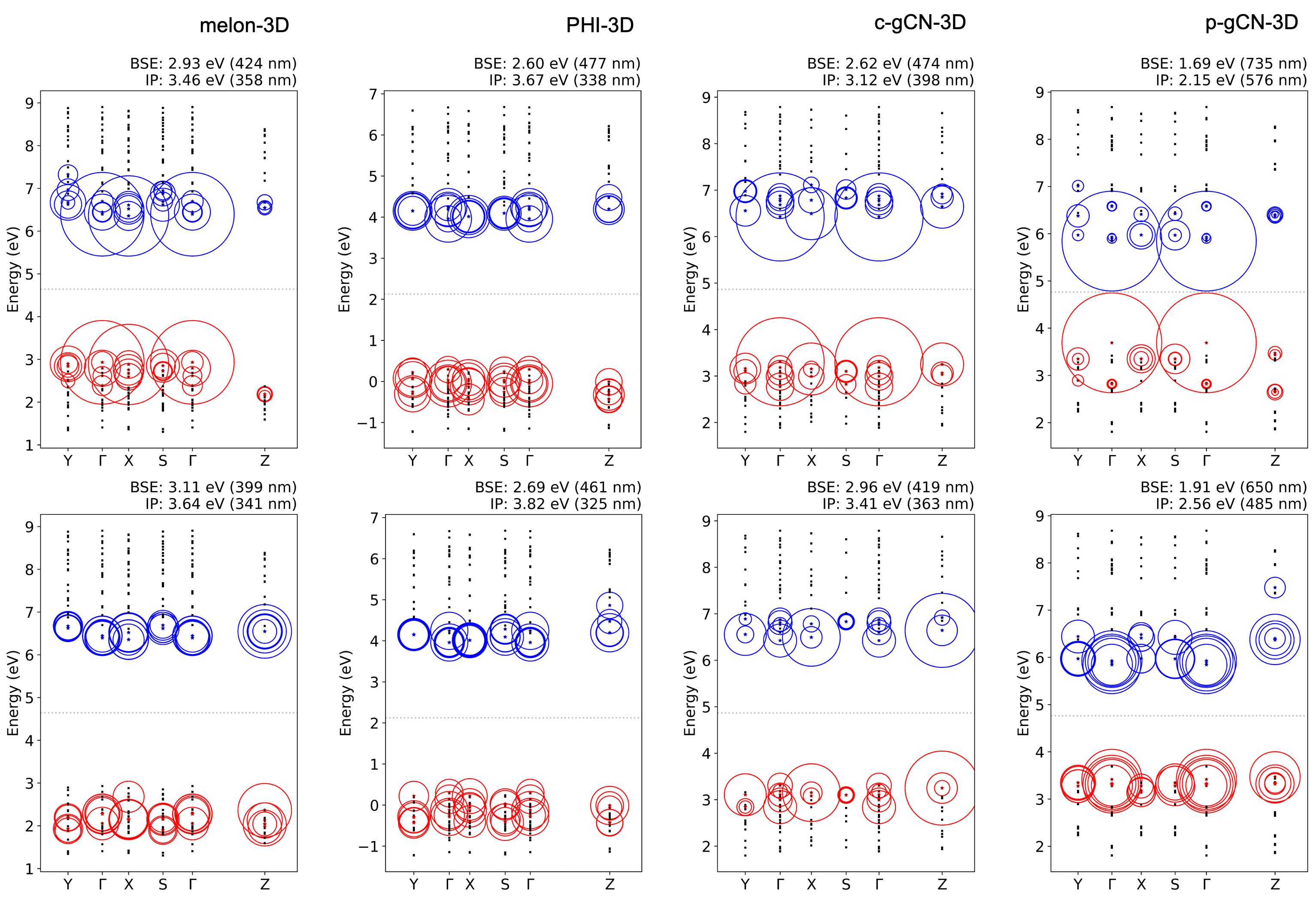}
    \caption{Representaton of eigenvalues (BSE and IP) and coupling coefficients (radius of circles)
    of bright states for both 2D and 3D PCN structures, corresponding to the absorption edges (top)
    and interlayer interactions (bottom).}
    \label{fig4}
\end{figure} 
\textit{Exciton formation}.---Figure\,\ref{fig3} demonstrates the oscillator strengths that, according
to the Franck--Condon principle, represent the probability of electronic transitions at each energy level. 
Cases of high oscillator strengths with relatively low absorption peaks occur in the absorption spectrum
due to consideration of the dipole momentum and selection rules. Therefore, by comparing absorption spectra
with oscillator strengths, we can predict the types of electronic transitions. Notably, most electronic
transitions near the absorption edge in PCN 3D structures are combined with the dark states. 
Our results from the BSE calculations provide the coupling coefficients corresponding to transitions between energy levels at each high symmetry point, as shown in Figure\,\ref{fig4}. The coupling constants of the other lowest 32 excited states are presented in the Supplemental Material. 
The coupling coefficients (radius of circles) at the band edges for each PCN 3D structure are displayed at the top line of the figure, and the coupling coefficients that are most pronounced for interlayer interactions are displayed at the bottom line. 
At the absorption edges, stable excitons are related to transitions between the band edges in melon, c-gCN, and p-gCN structures while relatively more degenerate excitons are formed in the PHI structure.
The coupling constants at the bottom line of Figure\,\ref{fig4} represent the exciton formation at the $Z$-point in each structure, indicating a direction toward the other layer from the $\Gamma$-point. 
These interlayer transitions are identified as the electron transitions from the sub-levels of the valence band to near the conduction band edge, indicating that these transitions involve different $\pi$ electrons than those associated with band edge transitions.
Our previous study found that absorption peaks of PCN structure can be sorted as in-plane $\pi$--$\pi^*$ transitions and $\pi$--$\pi^*$ interlayer transitions.\cite{im2024unraveling}
When comparing the relative stability of excitons formed in the PCN structures at the interlayer, the c-gCN structure exhibits the most stable exciton with an energy gap of 2.96\,eV. Melon and PHI also show stabilization of interlayer excitons at 3.11\,eV and 2.97\,eV, respectively, each showing roughly a 0.3 eV difference from their absorption edges.
The differences in exciton stability at the interlayer are determined by the degree of overlap between orbitals of different layers, specifically being regulated by the magnitude of the transition momentum off-diagonal components. 
These off-diagonal components turn out to be the degree of interaction of the nitrogen non-bonding ($n$) electrons, which impact the $\pi$ electrons at the interlayers. 
This explains how the interlayer distances are related to the configurations of PCN structures. 
This is more clearly understood when comparing the coupling constant plots of LDA-optimized geometries with
different interlayer distances (see Table\,S1) to those of GGA-optimized geometry results (Figs.\,S11, S13--S20). 
Furthermore, our previous study found that the thermodynamic contributions from vibrational motion
due to the flexible layered structure of PCNs are non-negligible.\cite{im2023structure} Thus, we expect that
enhanced interlayer interactions may be evident due to interlayer vibration modes.

\begin{figure}[hbp]
\centering
\includegraphics[width=0.80\textwidth]{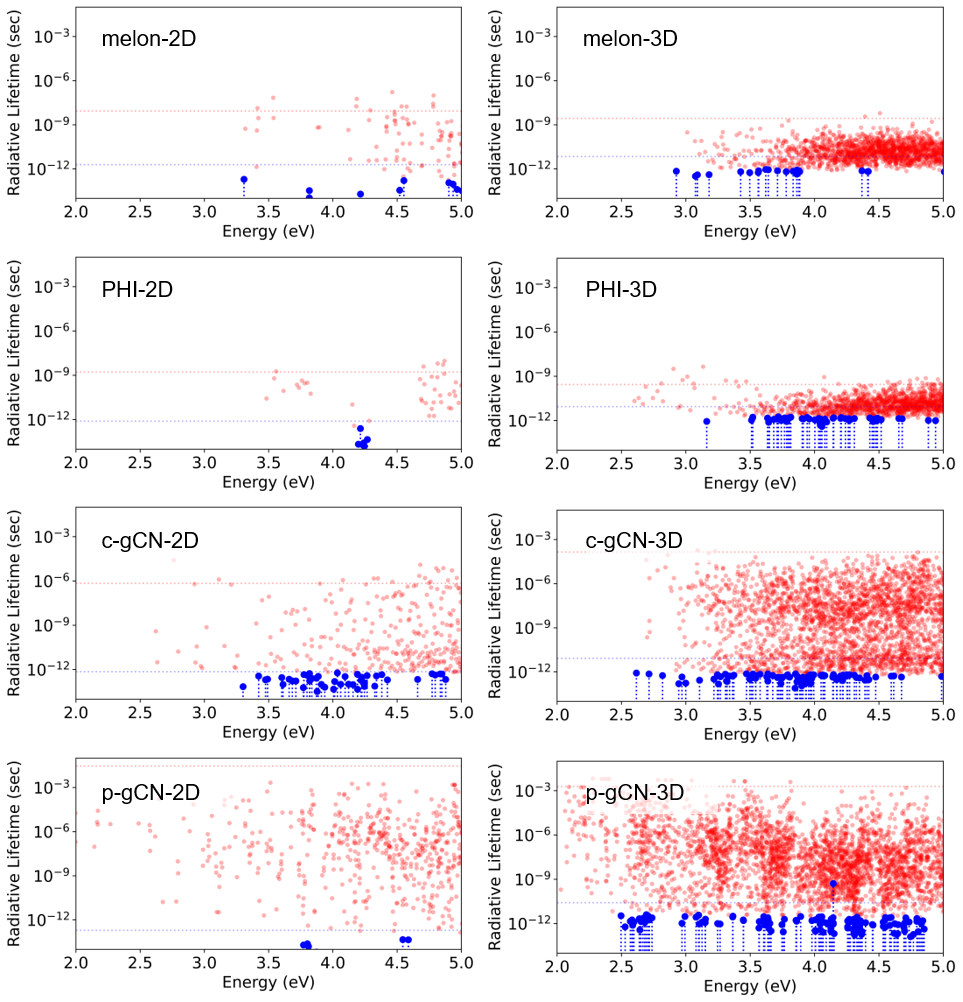}
    \caption{Calculated radiative exciton lifetimes of both 2D and 3D PCN structures. Bright states are
    denoted in blue, while dark states are represented in red. Their respective effective lifetimes at
    300\,K are indicated by dashed lines.}
    \label{fig5}
\end{figure}
\textit{Prolonged exciton lifetime and photocatalytic activity}.---We have calculated the exciton
lifetimes and effective lifetimes at 300\,K based on the energy required for each excited state,
transition probability, and transition dipole momentum. The obtained behavior is shown in Figure\,\ref{fig5}
and summarized in Table\,\ref{table1}. Here, we define the energy states with the top 5$\%$ oscillator
strengths as bright states (blue) and the rest as dark states (red) (see Method section). 
Firstly, the effective lifetimes of bright excitons in 3D PCN structures are slightly increased compared to those in 2D structures. 
This suggests that most bright excitons in PCN structures are formed by the direct in-plane transitions of strongly localized $\pi$ electrons and the interlayer $\pi$ transitions. Therefore, the stacking configuration helps stabilize these bright excitons. 
In contrast, the dark excitons formed in 3D structures are destabilized by interlayer interactions.
Most dark excitons are related to in-plane $n$--$\pi^*$ or distorted $\pi$--$\pi^*$ transitions and interlayer interactions cannot sufficiently stabilize these states due to the lack of interaction.
Notably, the effective lifetime of dark excitons increases dramatically in graphitic structures (c-gCN and p-gCN).
This indicates that a new form of excitons, different from those in melon and PHI structures, exists in gCN. We propose that these long-lived excitons are closely related to the trapped excitons reported in the literature\cite{godin2017time}.
Furthermore, we suggest that the calculated lifetimes of dark excitons shed light on the photo-charging phenomenon observed in ionic-PHI.\cite{lau2017dark,adler2021photodoping,kasap2016solar,yang2019electron,li2021ultrafast,schlomberg2019structural}
Notably, the photocharging phenomenon persists in the proton-exchanged structure (H-PHI), indicating that the phenomenon is more attributed to the high symmetry crystalline regions rather than the influence of existing cations. 
However, the principle and the nature of electron accumulation in PHIs have not been elucidated so far.
Our results demonstrate that the extended lifetime of excitons is observed in both 3D stacked and 2D graphitic structures (c-gCN and p-gCN), indicating that these long lifetimes are primarily due to structural features rather than interlayer interactions.
In contrast, the PHI structures exhibit the shortest dark exciton lifetime even shorter than those of melon structures.
This might explain the general experimental observation that the photo(electro)catalytic activity and photoinduced electron accumulation in ionic (PHI-type) PCNs are enabled only in the presence of effective electron donors, such as alcohols, that can quickly extract the hole from the short-lived excitons in PHI-type PCNs.\cite{lau2017dark,adler2021photodoping,kasap2016solar,yang2019electron,li2021ultrafast,pulignani2022rational}
\begin{figure*}[btp]
\centering
\includegraphics[width=0.75\textwidth]{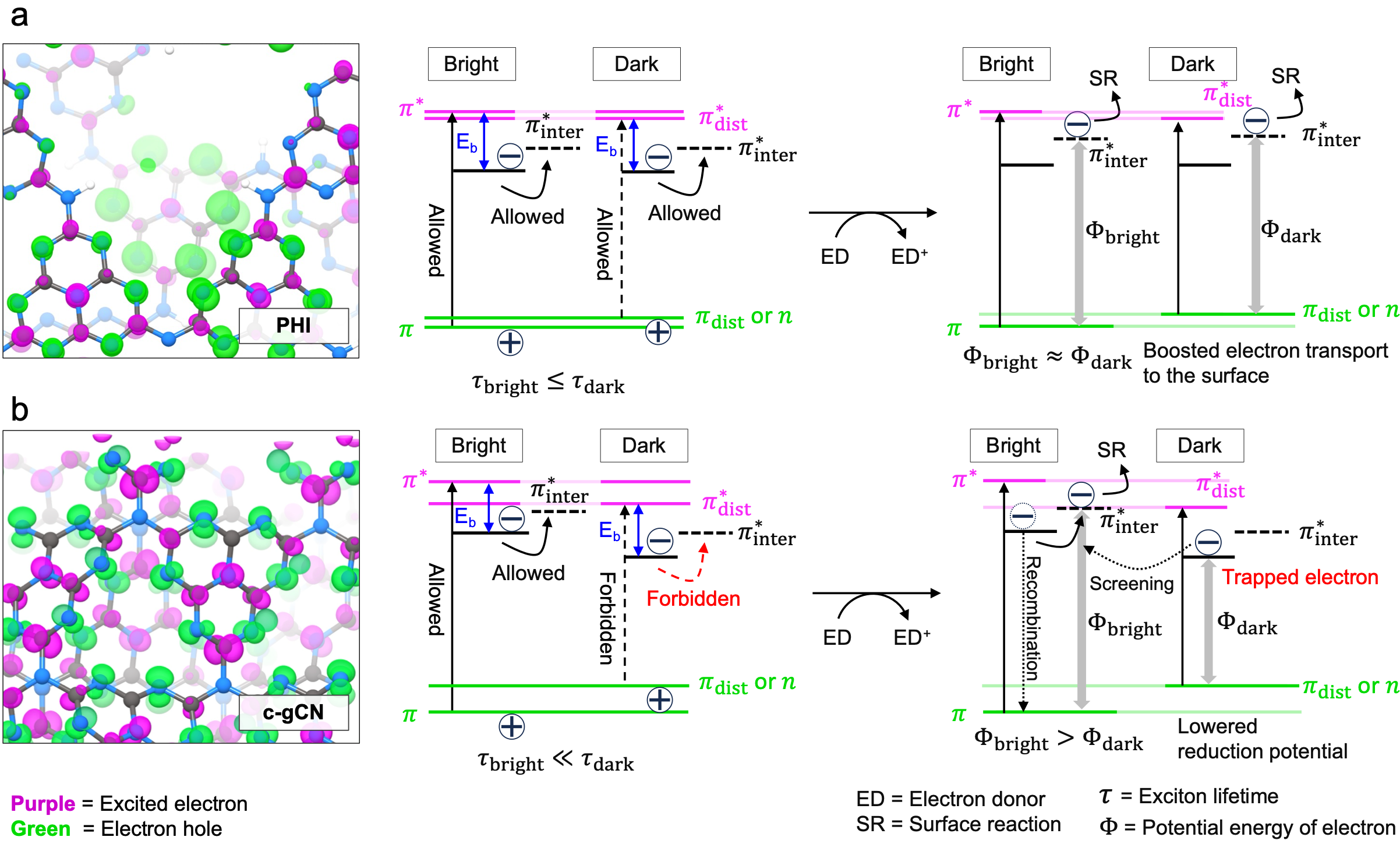}
    \caption{(left) Electron density of electron hole and the excited electrons and (right) the proposed mechanisms of photoexcitation process for (a) PHI and (b) c-gCN-3D structures, where $\pi_\mathrm{dist}$ and $\pi^*_\mathrm{dist}$ are the energy levels of distorted $\pi$ electrons, $\pi_\mathrm{inter}$ is the $\pi^*$ state at the other layer, and $\mathrm{E_b}$ is the exciton binding energy, respectively.}
\label{fig6}
\end{figure*}
In order to further elucidate the nature of these dark states, Figure\,\ref{fig6} shows the electronic density corresponding to the electron and the electron hole for the dark excitons of the longest lifetime in the PHI and the c-gCN structure.
It clearly shows that these dark excitons are different in PHI and c-gCN structures.
This suggests that the repulsive force between heptazine units causes a distorted orientation of the $n$ or $\pi$ orbitals.
The varying degrees of this distortion control the selection rules, resulting in allowed transitions in PHI and forbidden transitions in c-gCN for dark excitons.
This forbidden in-plane transition further hinders the interlayer interaction due to the lack of orbital overlaps.
Moreover, this distortion also affects the energy levels of $\pi$ states, resulting in a larger energy gap between $\pi^*$ and $\pi^*_\mathrm{dist}$ in c-gCN than in the case of PHI.
Therefore, the larger distortion of the orbitals can exhibit prolonged exciton lifetimes and lower energy gaps for the dark states. (See Table\,\ref{table1})
In the case of melon, the presence of hydrogen bonding slightly alters one of the $\pi$ electrons, resulting in dark states with lifetimes similar to but slightly longer than those in PHI.\cite{im2024unraveling}
In p-gCN, the strongly forced forbidden $n$–$\pi^*$ transitions due to the high symmetry lead to the detrimental effect of dark states, similar to those in c-gCN.
\begin{table}
\caption{Bandgaps of 2D and 3D PCN structures (optimized using CA functionals) were calculated via
         independent particle approximations ($G_0W_0$@IP) and BSE ($G_0W_0$@BSE) methods, alongside the
         effective exciton lifetime at 300\,K. (D) and (I) denote the direct and indirect
         bandgaps, respectively.}
\begin{tabular}{@{}ccccrr@{}}
\toprule
\multirow{2}{*}{} &
  \multirow{2}{*}{Structure} &
  \multirow{2}{*}{\begin{tabular}[c]{@{}c@{}}$G_0W_0$@IP\\      (eV)\end{tabular}} &
  \multirow{2}{*}{\begin{tabular}[c]{@{}c@{}}$G_0W_0$@BSE\\      (eV)\end{tabular}} &
  \multicolumn{2}{c}{\begin{tabular}[c]{@{}c@{}}Exiton lifetime\\ \textless{}$\tau_{\mathrm{300 K}}$\textgreater{}\end{tabular}} \\ \cmidrule(l){5-6} 
                    &       & \multicolumn{1}{l}{\textit{}} & \multicolumn{1}{l}{\textit{}} & bright   & dark\\ \midrule
\multirow{4}{*}{2D} & melon & 5.50 (D) & 3.31 & 1.96 ps  & 8.70 ns             \\
                    & PHI   & 5.83 (I) & 3.48 & 0.78 ps  & 1.65 ns            \\
                    & c-gCN    & 4.48 (D) & 2.62 & 8.36 ps  & \hspace*{0.3cm} 141.00 $\mu$s \\
                    & p-gCN    & 4.26 (I) & 1.95 & 0.20 ps  & 28.73 ms   \\ \midrule
\multirow{4}{*}{3D} & melon & 3.46 (I) & 2.93 & 6.96 ps  & 2.72 ns             \\
                    & PHI   & 3.67 (D) & 2.60 & 8.77 ps  & 277.34 ps           \\
                    & c-gCN    & 3.12 (D) & 2.62 & 8.36 ps  & 141.00 $\mu$s \\
                    & p-gCN    & 2.15 (I) & 1.69 & 26.04 ps & 1.92 ms    \\ \bottomrule
\end{tabular}
\label{table1}
\end{table} 
Furthermore, these findings shed light on the experimentally observed detrimental effects of deep charge trap states on photocatalytic activity in both conventional and ionic (PHI) carbon nitrides.\cite{godin2017time,yang2019electron}
The significant distortion of $\pi$ or $n$ orbitals can create deeply trapped dark excitons compared to the bright states (Fig.\,\ref{fig6}b).
Due to their longer exciton lifetimes, these trapped excitons are less affected by hole-quenching rates and can effectively accumulate electrons.
However, these electrons, being spatially distorted and trapped, can not directly contribute to the interlayer interactions for the carrier transport, resulting in significantly reduced quantum yields due to their strongly diminished reductive potential, as reported in the literature.\cite{godin2017time}
Moreover, the presence of these trapped electrons might shield interlayer charge transport of the bright states, which could further promote the recombination of bright excitons.
The presence of these deeply trapped electrons also suggests that they can be used as a descriptor to indicate whether conventional or ionic (PHI) carbon nitrides obtained from the synthesis process contain graphitic domains.
On the other hand, in PHI structures (Fig.\,\ref{fig6}a), the dark states of the less affected $\pi$ electrons exhibit similar energy levels to the bright states, forming shallow trapped states.
Due to this similarity, these dark states can have lifetimes as short as the bright states.
Although a fast electron donor rate is required for hole-quenching due to these short lifetimes, once the hole of the dark exciton is filled, the nearly undistorted orientation of the $\pi$ electrons enables efficient interlayer carrier transport, leading to effective surface catalytic reactions.
We assume that these dark-state-induced phenomena (enhanced photocatalytic activity or time-delayed photocatalysis) in PHI can be more pronounced in the presence of cations, as the stabilization of these excess electrons is also a key factor in preventing recombination.\cite{lau2017dark,adler2021photodoping,schlomberg2019structural}
We note that, although the model structures used in our simulations cannot fully account for the complex and still not fully resolved structure of real-life PCN materials, our results are in excellent agreement with experimental observations, and elucidate the crucial role of dark states in photocatalytic activity of PCN materials.

\textit{Conclusion}.---Our study elucidates the principles underlying the exciton formation, transitions,
and lifetimes within PCN structures. The results establish a valuable link between the structural and optical properties of PCN materials, which is a key prerequisite for rational tuning of their electronic structures for optimal performance.
Utilizing \textit{GW}@BSE calculations, we find the decisive role of dark excitons for effective photocatalytic performance. The structure-dependent mechanisms of dark exciton formation elucidate the electron accumulation phenomena and subsequent surface reactions in different types of carbon nitrides. Our study thus establishes a valuable theoretical framework for further spectroscopic studies, and paves the way for the knowledge-driven design of PCN-based materials with tuned properties. \newline

This work was funded by the Deutsche Forschungsgemeinschaft (DFG -- German Research Foundation) through
TRR-234 CataLight (project no. 364549901) as well as JA 1072/27-1 and BE 5102/5-1 (project no. 428764269).
R.B. acknowledges funding from the European Union's Horizon Europe programme for research and innovation
under grant agreement No. 101122061 (SUNGATE). The authors acknowledge support by the state of 
Baden-W\"urttemberg through bwHPC and the German Research Foundation (DFG) through grant no INST 40/575-1
FUGG (JUSTUS 2 cluster). C.I. acknowledges the German Academic Exchange Service (DAAD, Ref. No. 91676720). 
\newpage
\bibliography{apssamp}
\setcounter{figure}{0}
\setcounter{table}{0}
\renewcommand{\thefigure}{S\arabic{figure}}
\renewcommand{\thetable}{S\arabic{table}}
\section{Supplementary data}

\begin{table}[h]
\caption{The lattice parameters of the PCN structures obtained by the CAPZ and PBE functionals.}
\begin{tabular}{@{}ccccccccc@{}}
\toprule
\multirow{2}{*}{} &
  \multicolumn{1}{l}{\multirow{2}{*}{}} &
  \multicolumn{3}{c}{\begin{tabular}[c]{@{}c@{}}Lattice parameters\\ (CA) (\r{A})\end{tabular}} &
   &
  \multicolumn{3}{c}{\begin{tabular}[c]{@{}c@{}}Lattice parameters\\  (PBE) (\r{A})\end{tabular}} \\ \cmidrule(r){3-5} \cmidrule(l){7-9}
                    & \multicolumn{1}{l}{} & \textit{a} & \textit{b} & \textit{c} &  & \textit{a} & \textit{b} & \textit{c} \\ \midrule 
\multirow{4}{*}{2D} & melon   & 16.45      & 12.57      & 12         &  & 16.42      & 12.62      & 12         \\
                    & PHI     & 22.29      & 12.82      & 12         &  & 21.58      & 12.46      & 12         \\
                    & c-gCN   & 13.28      & 11.8       & 12         &  & 13.35      & 11.88      & 12         \\
                    & p-gCN   & 14.15      & 14.15      & 12         &  & 14.23      & 14.23      & 12         \\ \midrule  
\multirow{4}{*}{3D} & melon   & 16.29      & 12.45      & 5.91       &  & 16.42      & 12.62      & 6.53       \\
                    & PHI     & 21.71      & 12.55      & 5.6        &  & 21.58      & 12.46      & 6.51       \\
                    & c-gCN   & 13.14      & 11.81      & 6.5        &  & 13.29      & 11.84      & 7.35       \\
                    & p-gCN   & 14.13      & 14.13      & 6.04       &  & 14.23      & 14.23      & 6.66       \\ \bottomrule
\end{tabular}
\end{table}

\begin{table}[h]
\caption{Bandgaps of 2D and 3D PCN structures (optimized using PBE functionals) were calculated via independent particle approximations ($G_0W_0$@IP) and BSE ($G_0W_0$@BSE) methods, alongside the effective exciton lifetime at 300 K. The (D) and (I) denote the direct and the indirect bandgap, respectively.}
\begin{tabular}{@{}ccccrr@{}}
\toprule
\multirow{2}{*}{} &
  \multirow{2}{*}{Structure} &
  \multirow{2}{*}{\begin{tabular}[c]{@{}c@{}}$G_0W_0$@IP\\      (eV)\end{tabular}} &
  \multirow{2}{*}{\begin{tabular}[c]{@{}c@{}}$G_0W_0$@BSE\\      (eV)\end{tabular}} &
  \multicolumn{2}{c}{\begin{tabular}[c]{@{}c@{}}Exiton lifetime\\ \textless{}$\tau_{\mathrm{300 K}}$\textgreater{}\end{tabular}} \\ \cmidrule(l){5-6} 
                    &       & \multicolumn{1}{l}{\textit{}} & \multicolumn{1}{l}{\textit{}} & bright   & dark\\ \midrule
\multirow{4}{*}{2D} & melon & 5.44 (D) & 3.24 & 1.87 ps  & \hspace*{0.3cm}196.36 ps           \\
                    & PHI   & 5.26 (I) & 2.03 & 5.35 ps  & 633.69 ps           \\
                    & c-gCN & 4.47 (D) & 2.63 & 5.18 ps  & 63.82 ns            \\
                    & p-gCN & 4.27 (D) & 1.97 & 0.16 ps  & 747.08 $\mu$s \\ \midrule
\multirow{4}{*}{3D} & melon & 3.82 (D) & 3.22 & 4.61 ps  & 23.29 ns            \\
                    & PHI   & 4.25 (I) & 2.53 & 20.34 ps & 289.93 ps           \\
                    & c-gCN & 3.39 (D) & 2.67 & 3.32 ps  & 2.88 ns             \\
                    & p-gCN & 2.87 (I) & 2.00 & 8.24 ps  & 31.49 ms   \\ \bottomrule
\end{tabular}
\end{table}
\clearpage

\begin{figure}[h]
\centering
\includegraphics[width=0.9\textwidth]{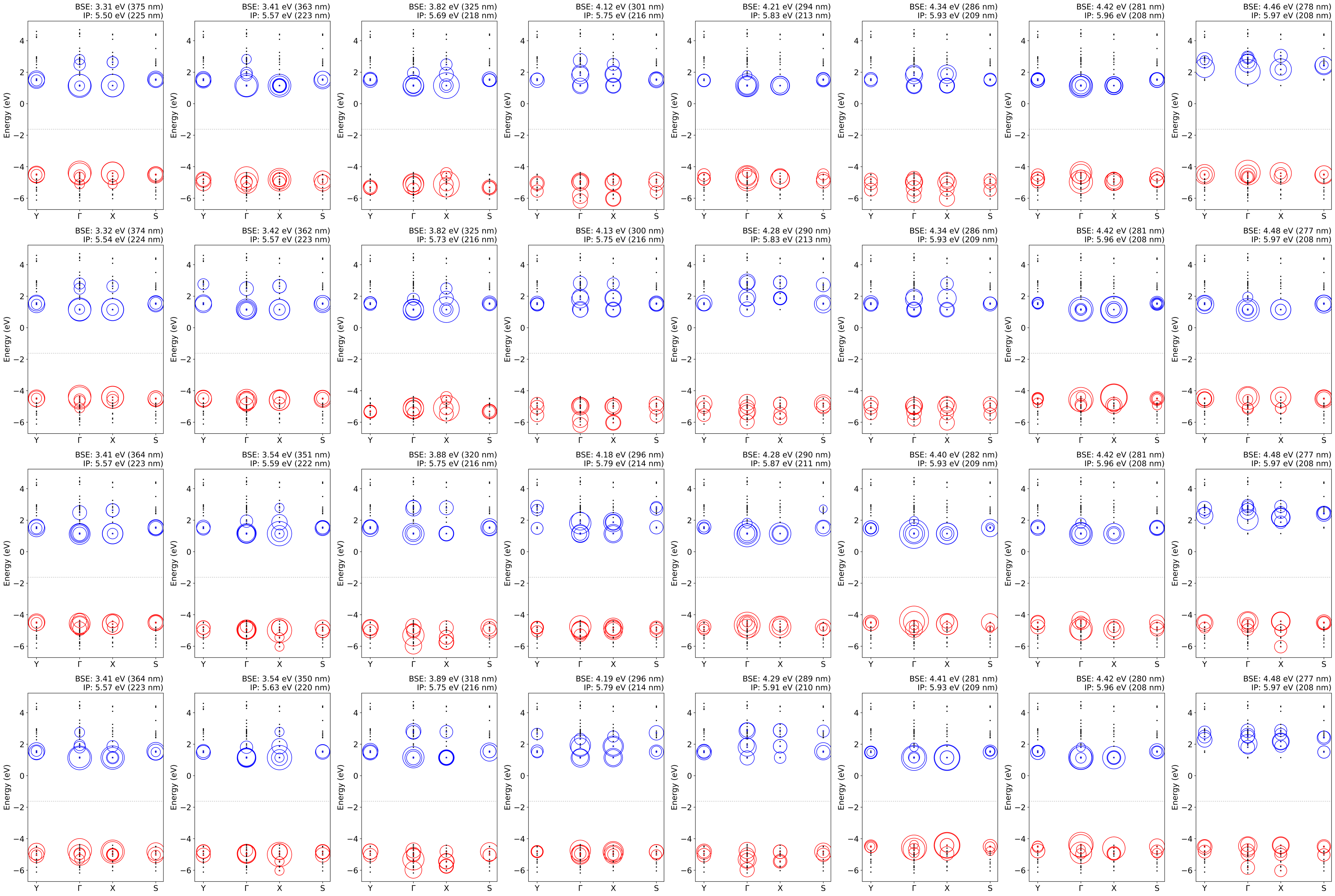}
\caption{Illustration of the $e$--$h$ coupling constants corresponding to the 32 excitation energy of melon-2D structure optimized CA (LDA) functionals.}
\end{figure}

\begin{figure}
\centering
\includegraphics[width=0.9\textwidth]{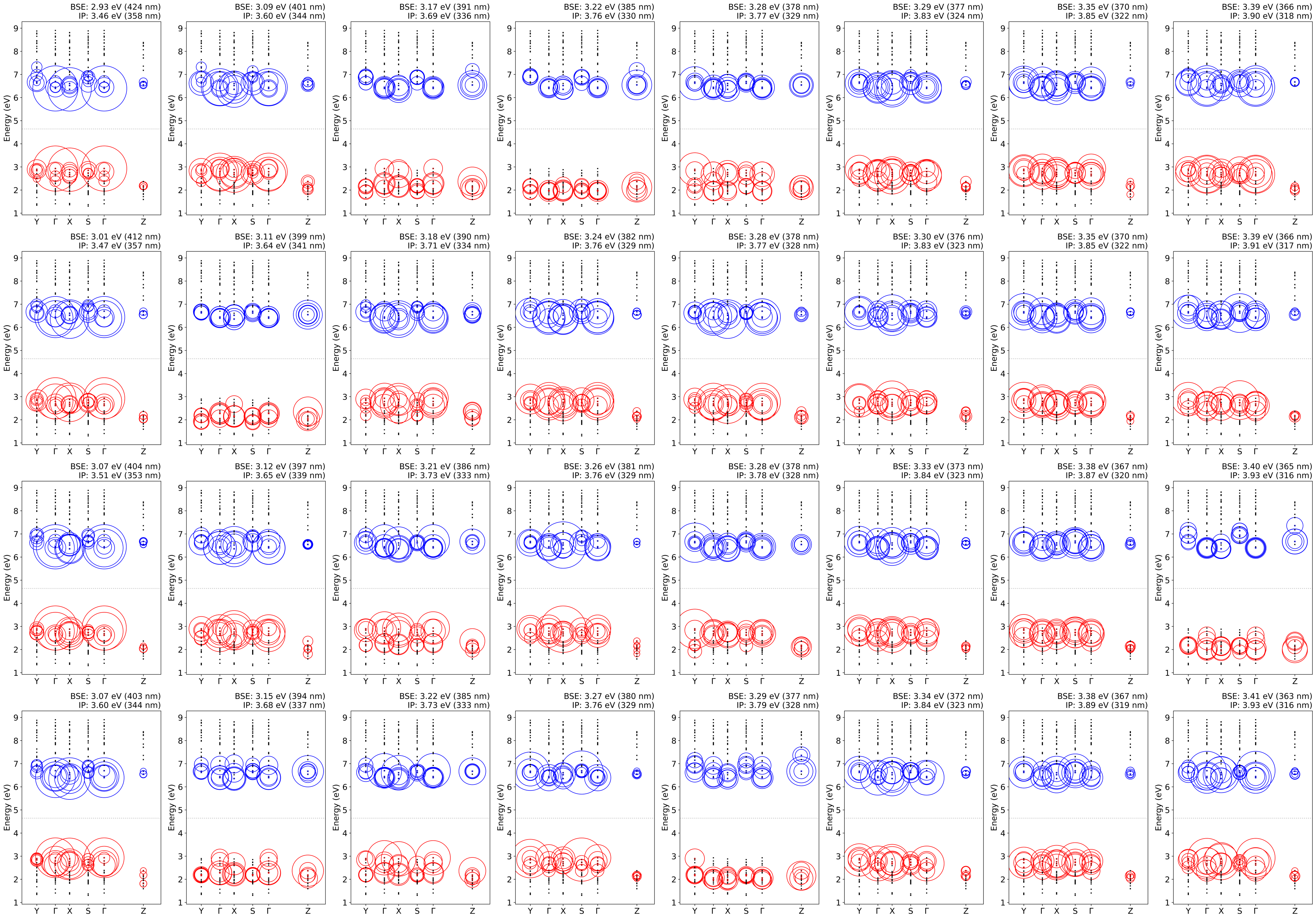}
\caption{Illustration of the $e$--$h$ coupling constants corresponding to the 32 excitation energy of melon-3D structure optimized CA (LDA) functionals.}
\end{figure}

\begin{figure}
\centering
\includegraphics[width=0.9\textwidth]{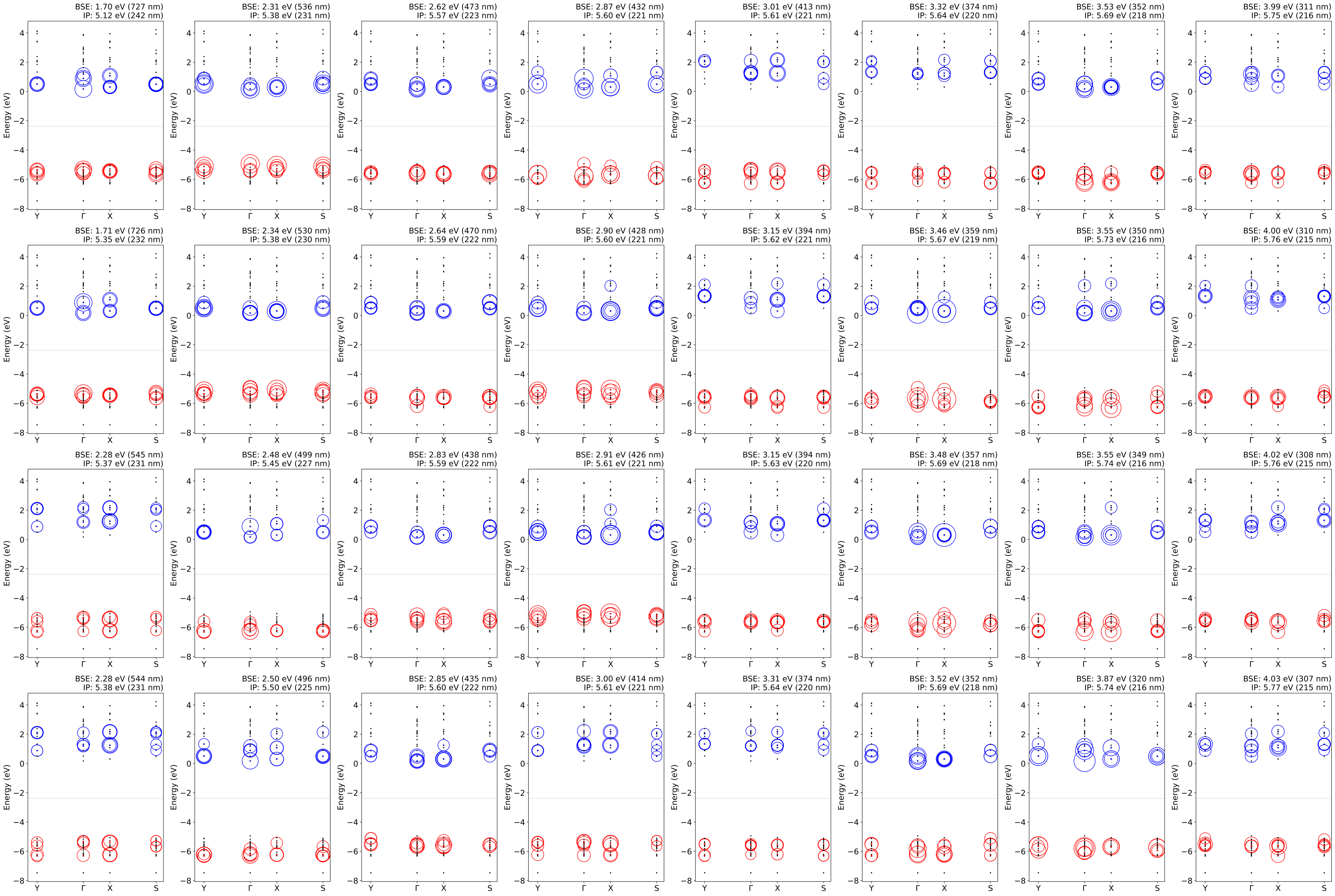}
\caption{Illustration of the $e$--$h$ coupling constants corresponding to the 32 excitation energy of PHI-2D structure optimized CA (LDA) functionals.}
\end{figure}

\begin{figure}
\centering
\includegraphics[width=0.9\textwidth]{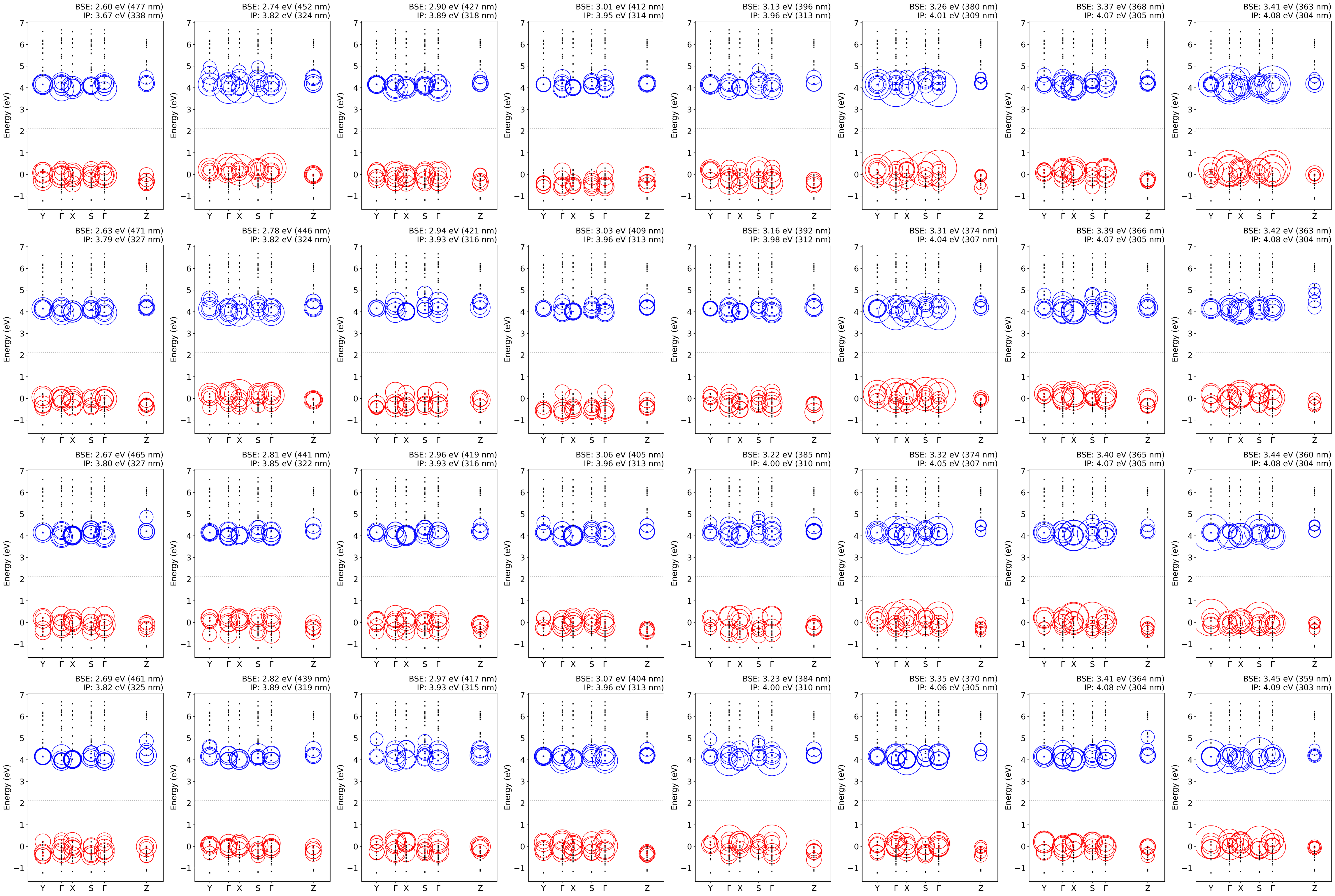}
\caption{Illustration of the $e$--$h$ coupling constants corresponding to the 32 excitation energy of PHI-3D structure optimized CA (LDA) functionals.}
\end{figure}

\begin{figure}
\centering
\includegraphics[width=0.9\textwidth]{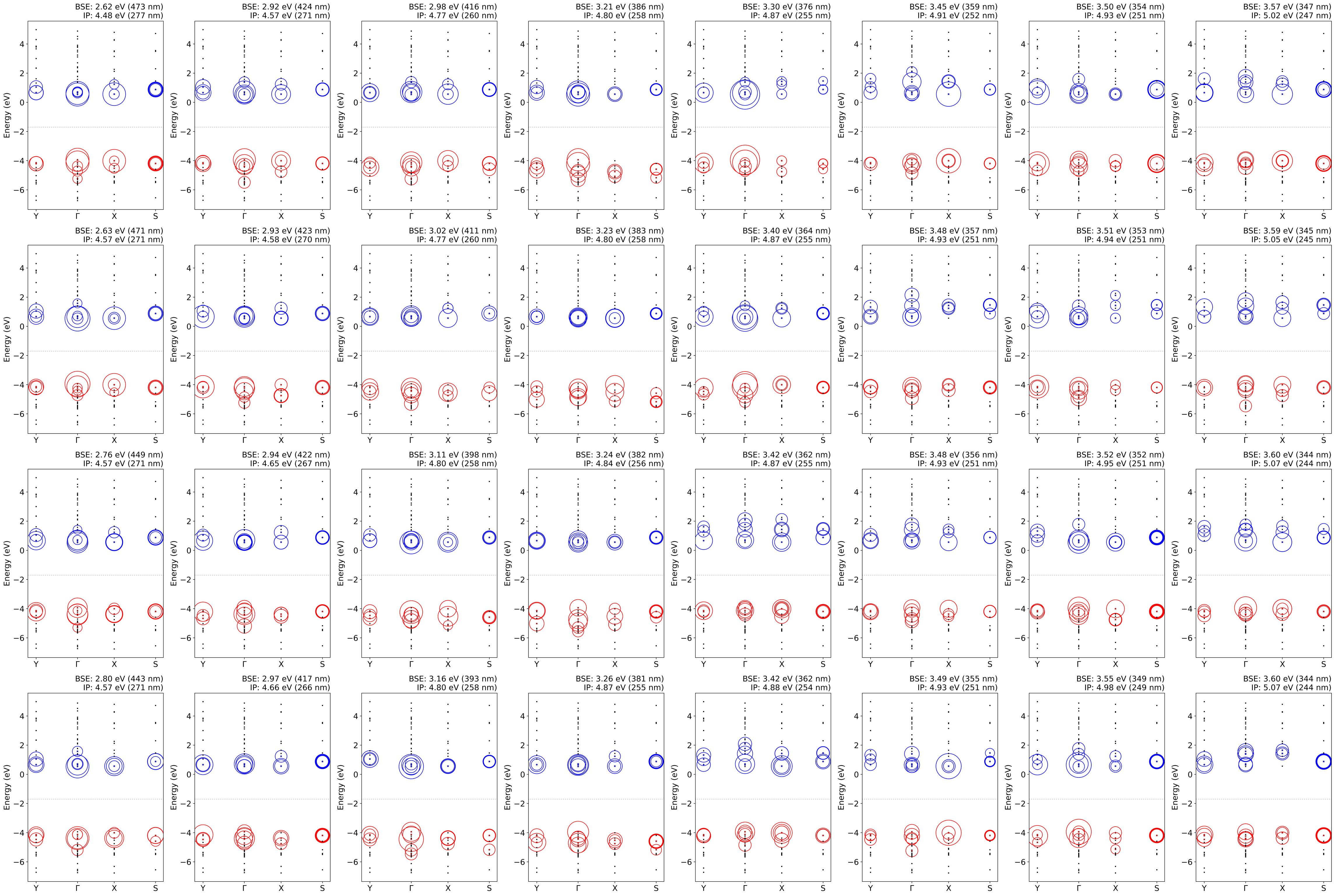}
\caption{Illustration of the $e$--$h$ coupling constants corresponding to the 32 excitation energy of c-gCN-2D structure optimized CA (LDA) functionals.}
\end{figure}

\begin{figure}
\centering
\includegraphics[width=0.9\textwidth]{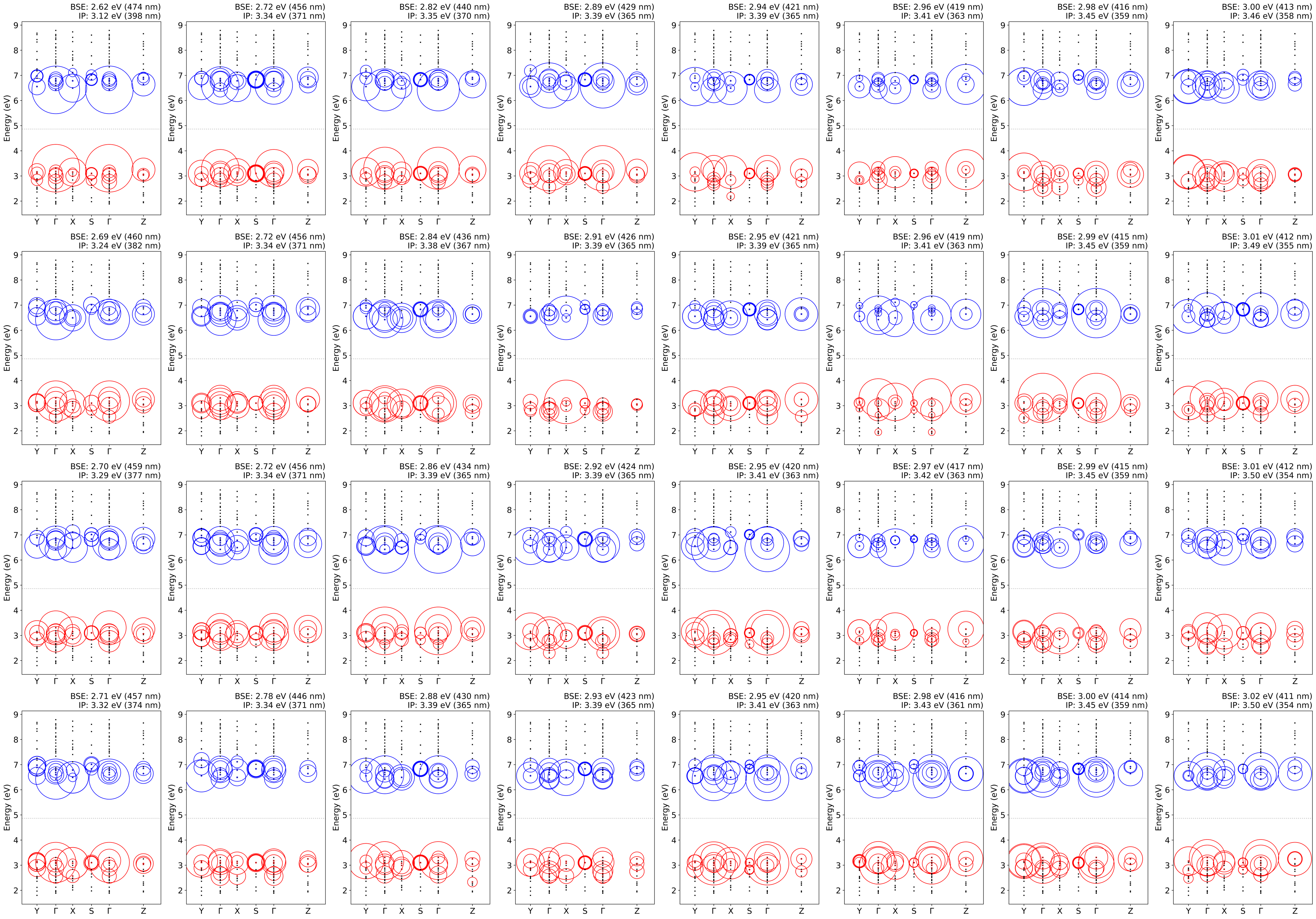}
\caption{Illustration of the $e$--$h$ coupling constants corresponding to the 32 excitation energy of c-gCN-3D structure optimized CA (LDA) functionals.}
\end{figure}

\begin{figure}
\centering
\includegraphics[width=0.9\textwidth]{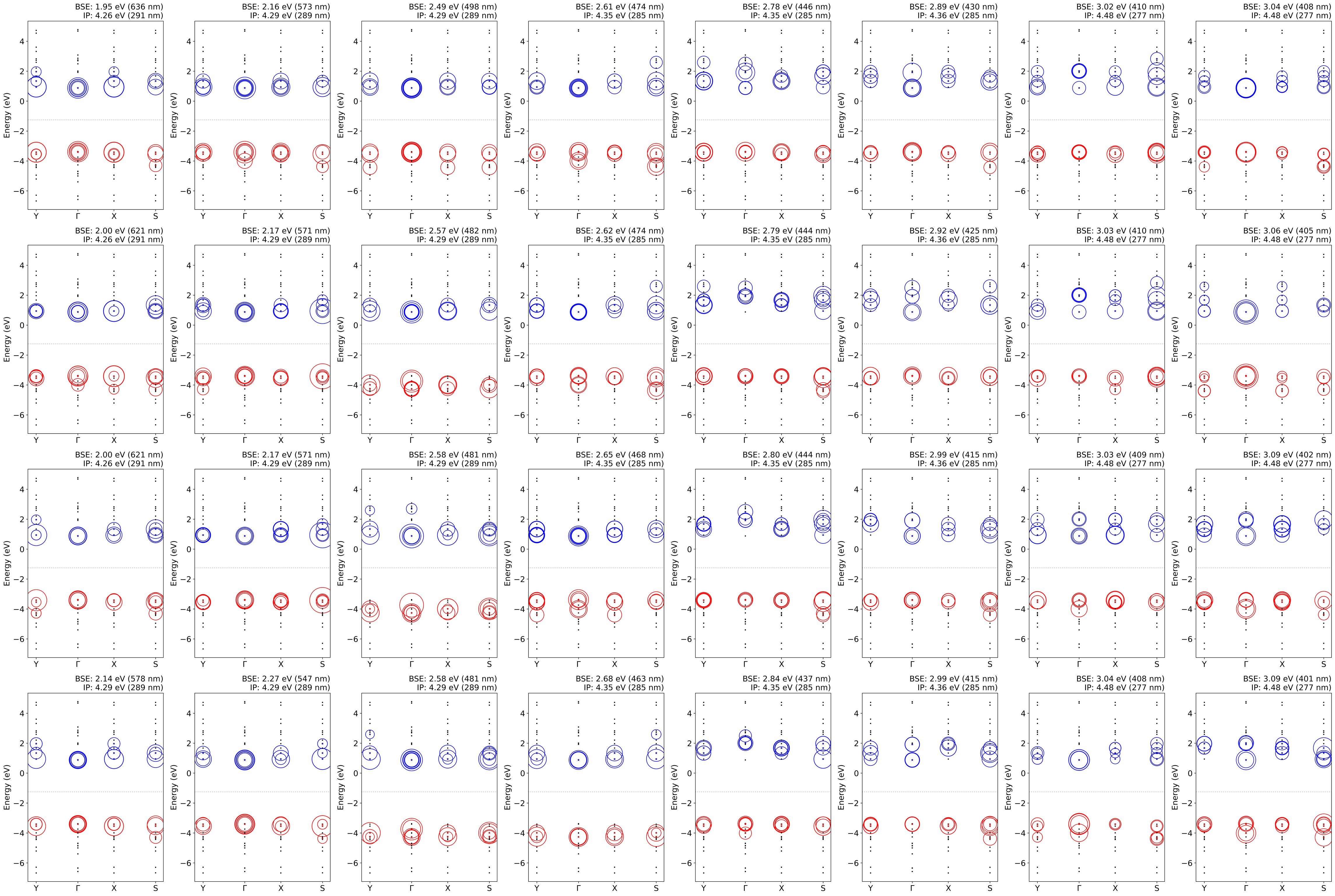}
\caption{Illustration of the $e$--$h$ coupling constants corresponding to the 32 excitation energy of p-gCN-2D structure optimized CA (LDA) functionals.}
\end{figure}

\begin{figure}
\centering
\includegraphics[width=0.9\textwidth]{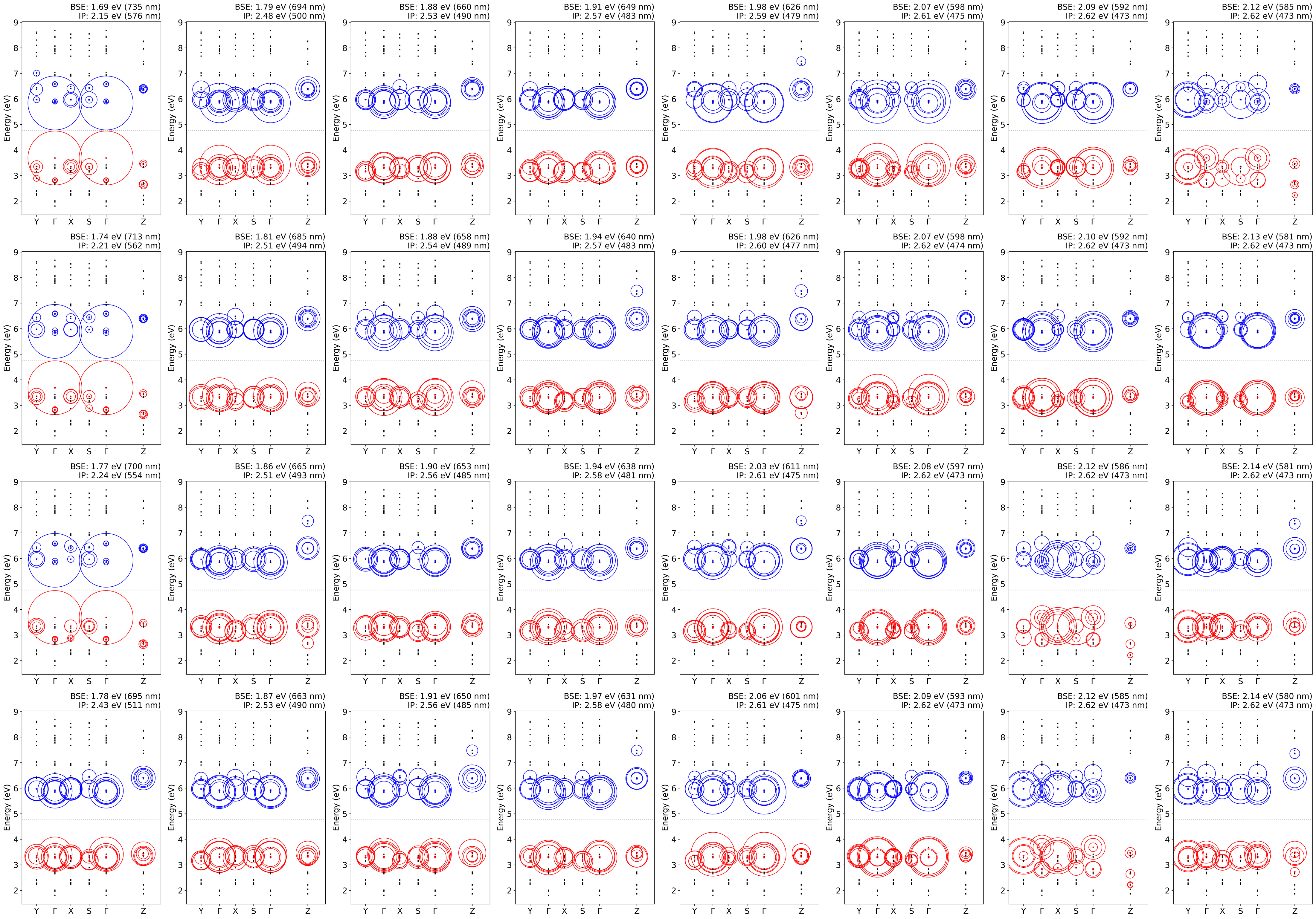}
\caption{Illustration of the $e$--$h$ coupling constants corresponding to the 32 excitation energy of p-gCN-3D structure optimized CA (LDA) functionals.}
\end{figure}

\newpage

\begin{figure}
\centering
\includegraphics[width=\textwidth]{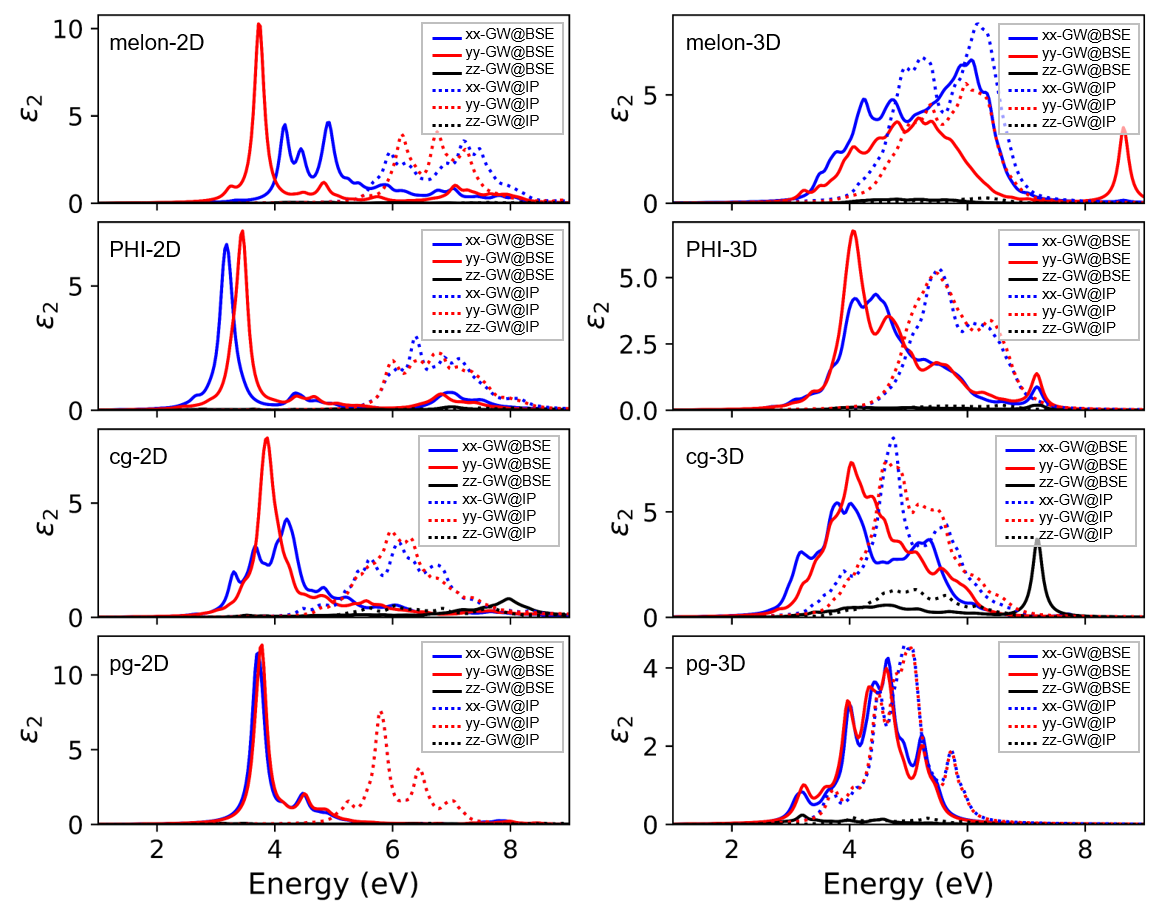}
\caption{Imaginary part of dielectric functions of the PCN structure optimized with PBE-D3 functionals. The solid and the dashed lines are obtained by the \textit{GW}@BSE and the \textit{GW}@IP respectively.}
\end{figure}

\newpage

\begin{figure}
\centering
\includegraphics[width=\textwidth]{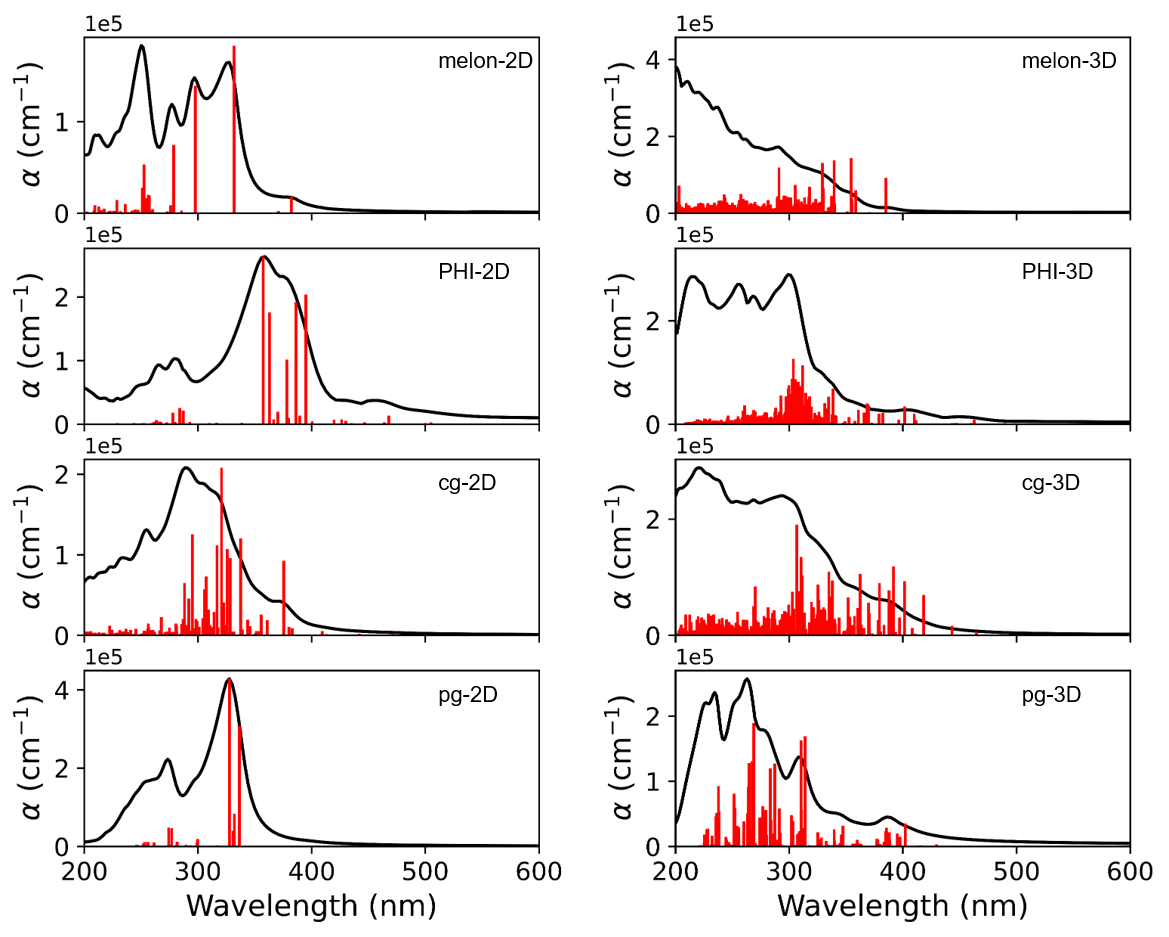}
\caption{The illustration of the calculated absorption spectra for 2D and 3D PCN models, optimized PBE functionals, along averaged spatial directions. The oscillator strengths in relation to wavelength are presented in red.}
\end{figure}

\newpage

\begin{figure}
\centering
\includegraphics[width=\textwidth]{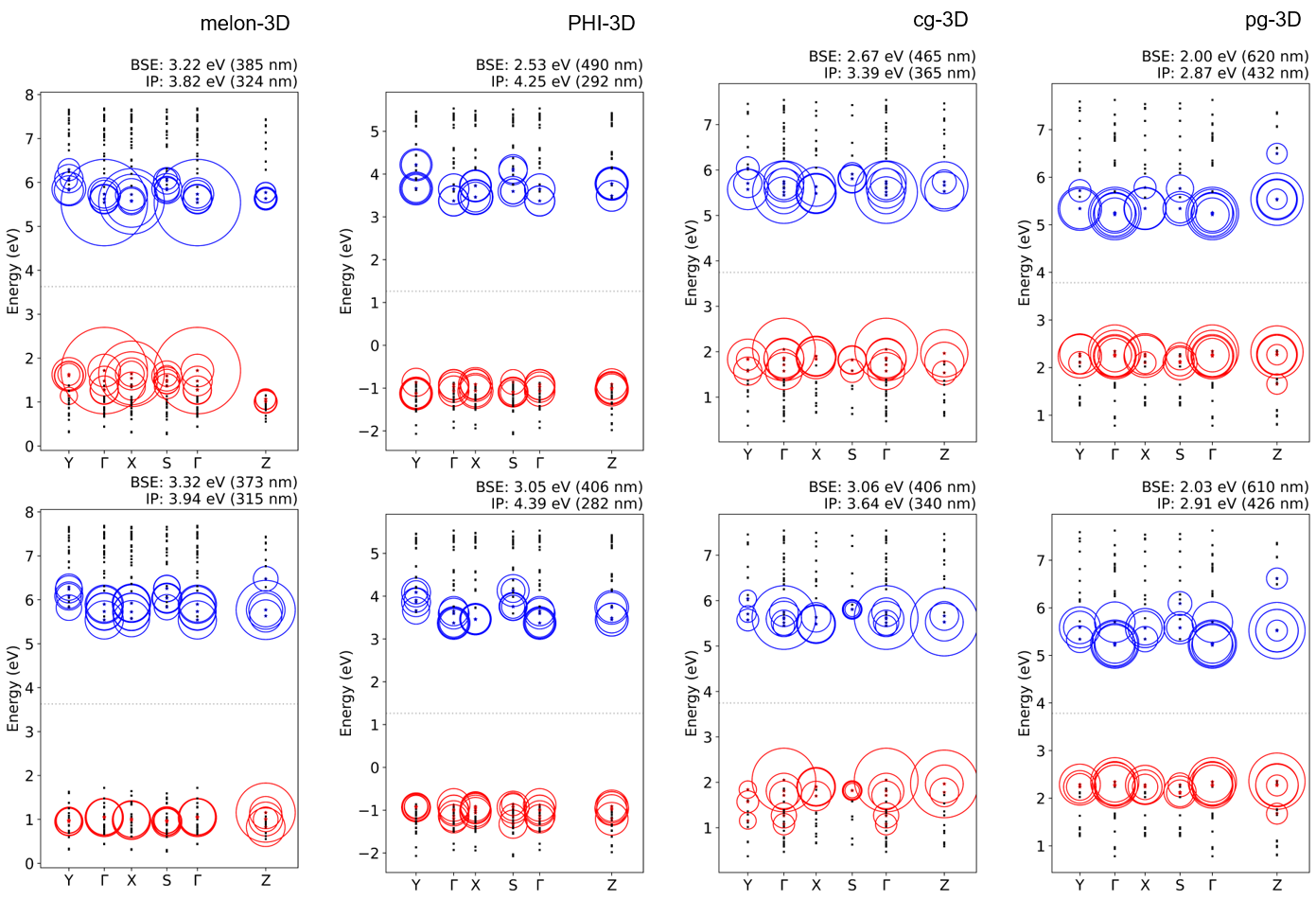}
\caption{Representaton of eigenvalues (BSE and IP) and coupling coefficients (radius of circles) of bright states for both 2D and 3D PCN structures, optimized by PBE-D3 functionals, corresponding to the absorption edges (top) and interlayer interactions (bottom).}
\end{figure}

\newpage

\begin{figure}
\centering
\includegraphics[width=\textwidth]{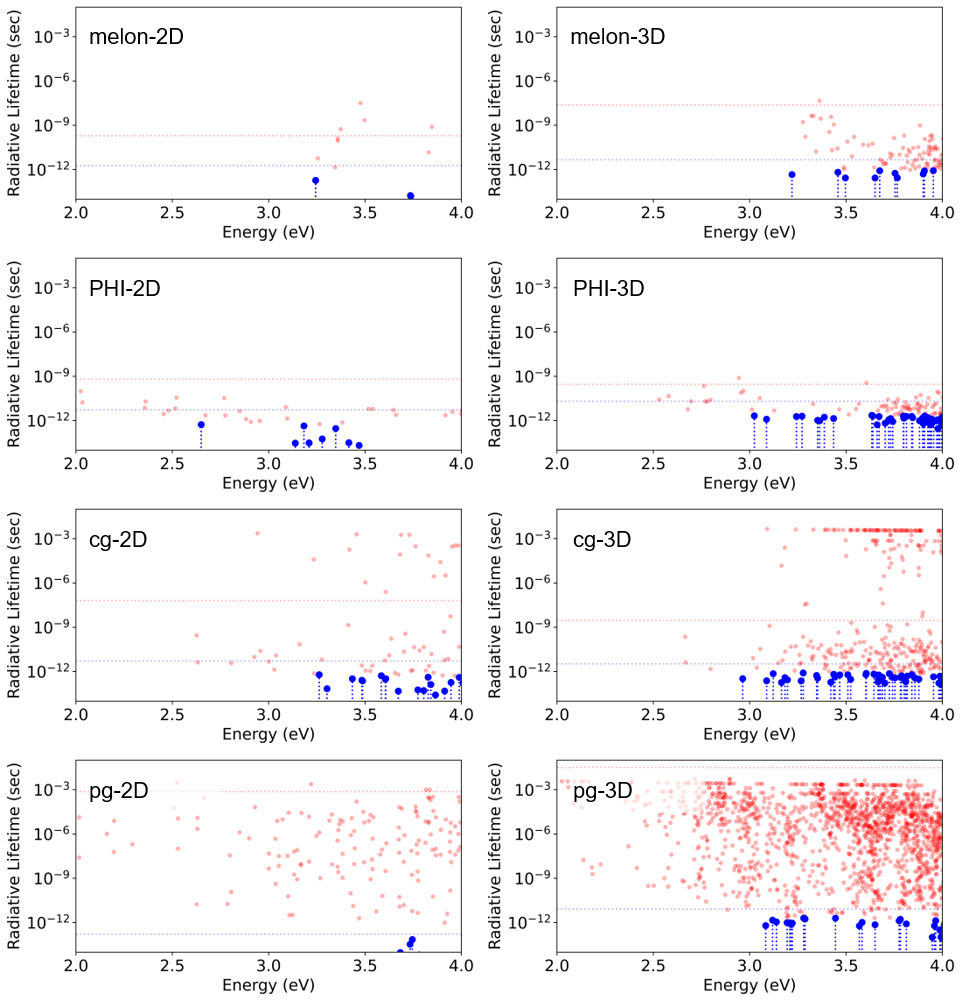}
\caption{Calculated radiative exciton lifetime of both 2D and 3D PCN structures optimized by PBE-D3 functionals. Bright states are denoted in blue, while dark states are represented in red. Their respective effective lifetime at 300 K are indicated by dashed lines.}
\end{figure}

\newpage

\begin{figure}
\centering
\includegraphics[width=0.9\textwidth]{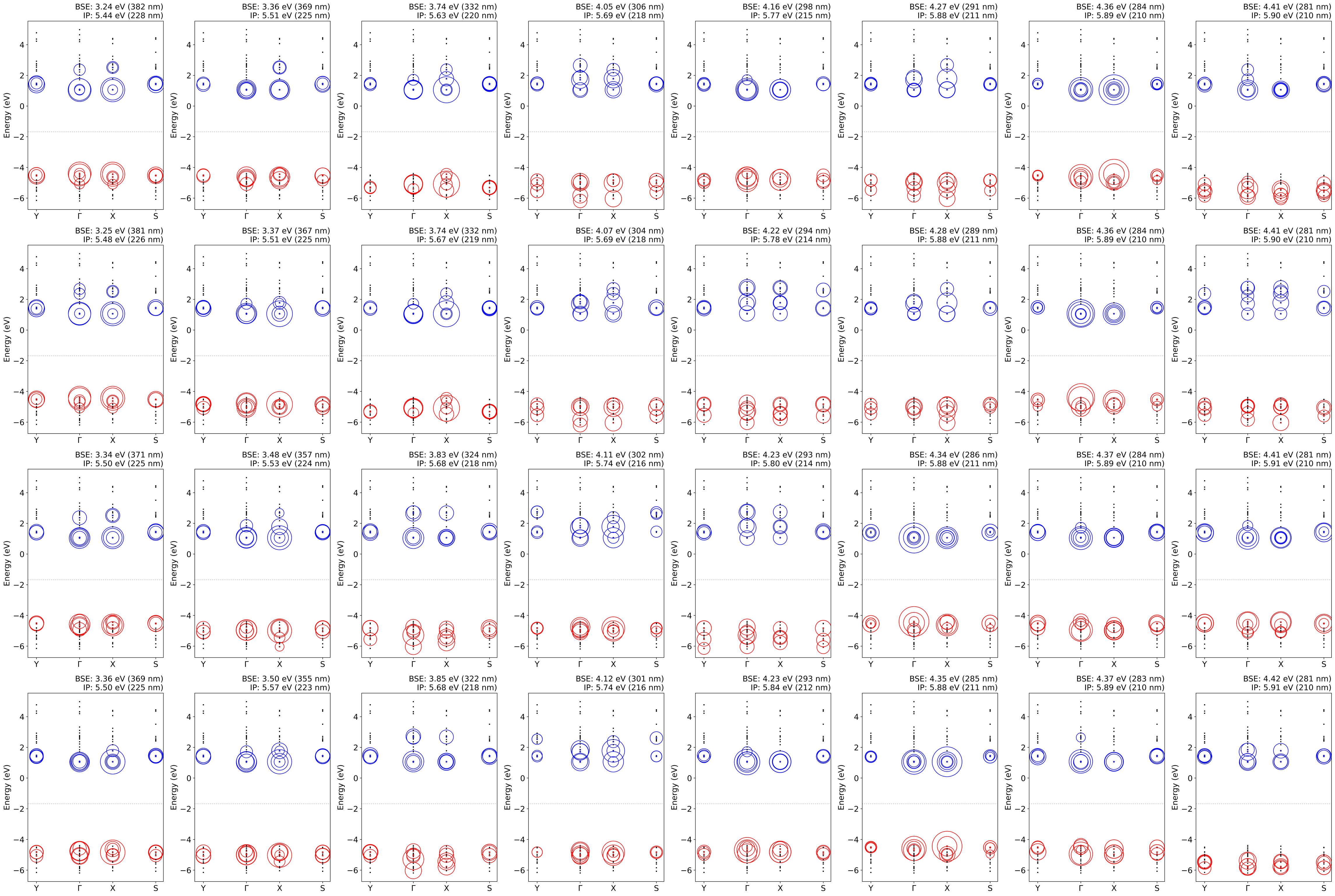}
\caption{Illustration of the $e$--$h$ coupling constants corresponding to the 32 excitation energy of melon-2D structure optimized PBE (GGA) functionals.}
\end{figure}
\newpage

\begin{figure}
\centering
\includegraphics[width=0.9\textwidth]{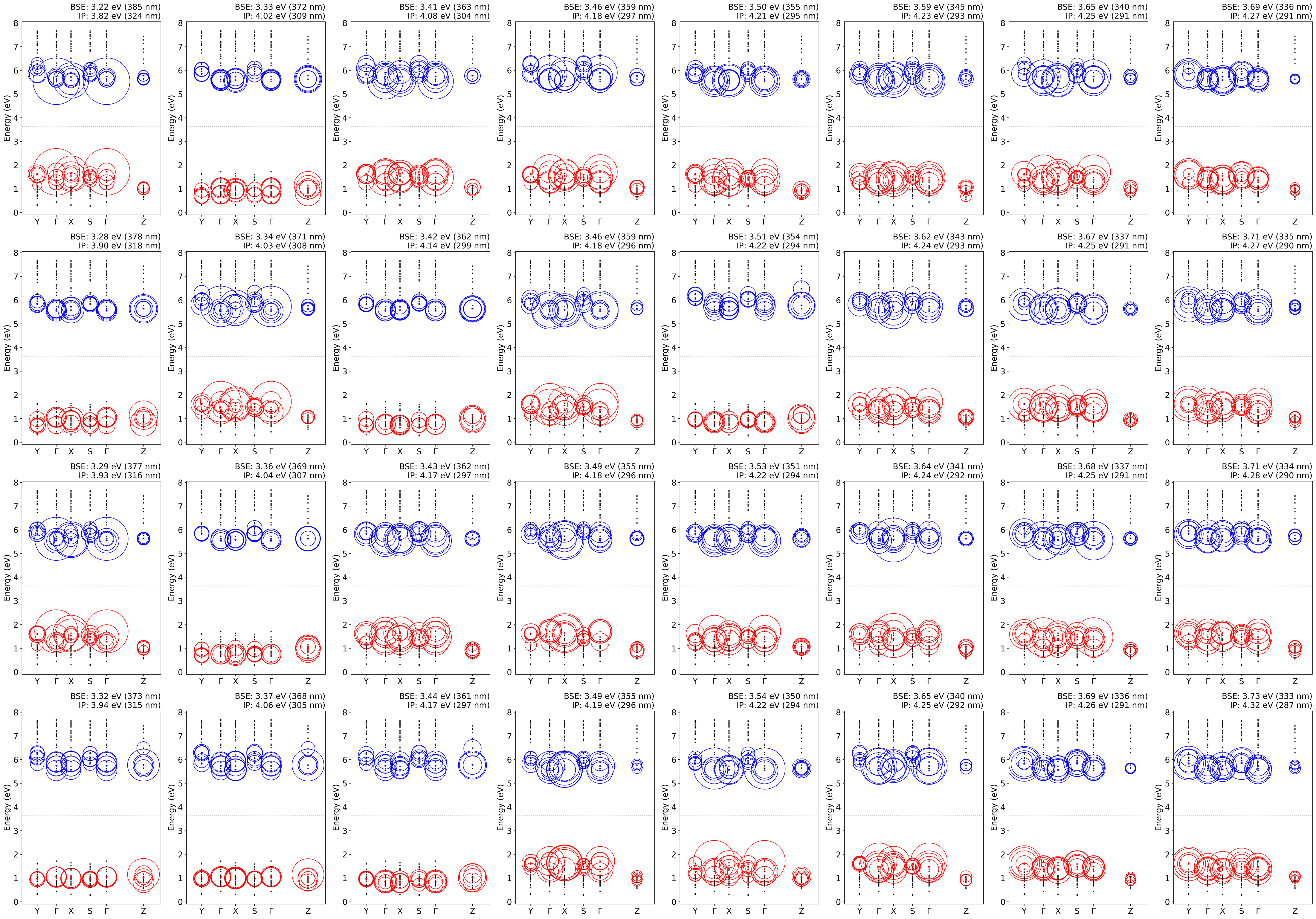}
\caption{Illustration of the $e$--$h$ coupling constants corresponding to the 32 excitation energy of melon-3D structure optimized PBE (GGA) functionals.}
\end{figure}
\newpage

\begin{figure}
\centering
\includegraphics[width=0.9\textwidth]{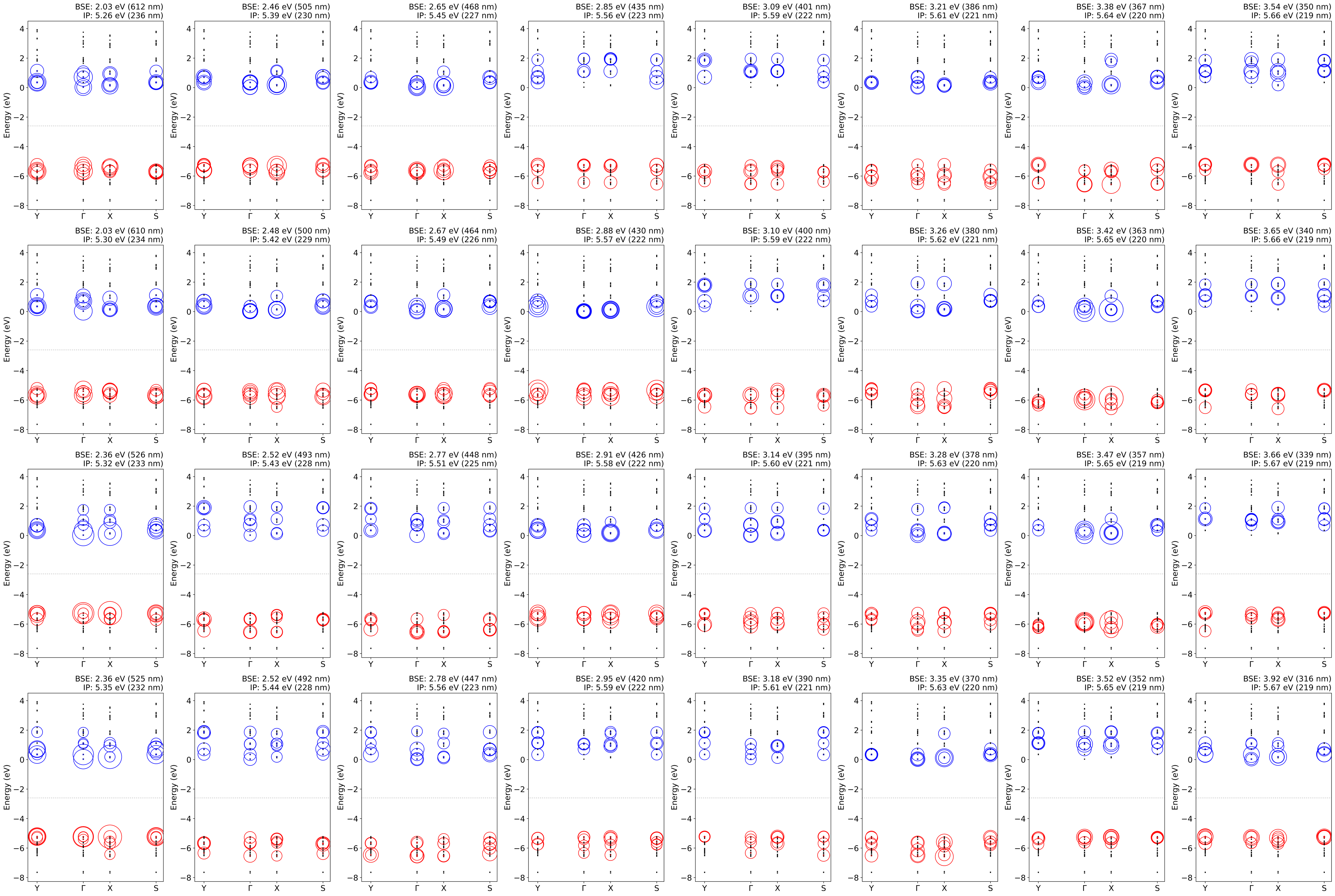}
\caption{Illustration of the $e$--$h$ coupling constants corresponding to the 32 excitation energy of PHI-2D structure optimized PBE (GGA) functionals.}
\end{figure}
\newpage

\begin{figure}
\centering
\includegraphics[width=0.9\textwidth]{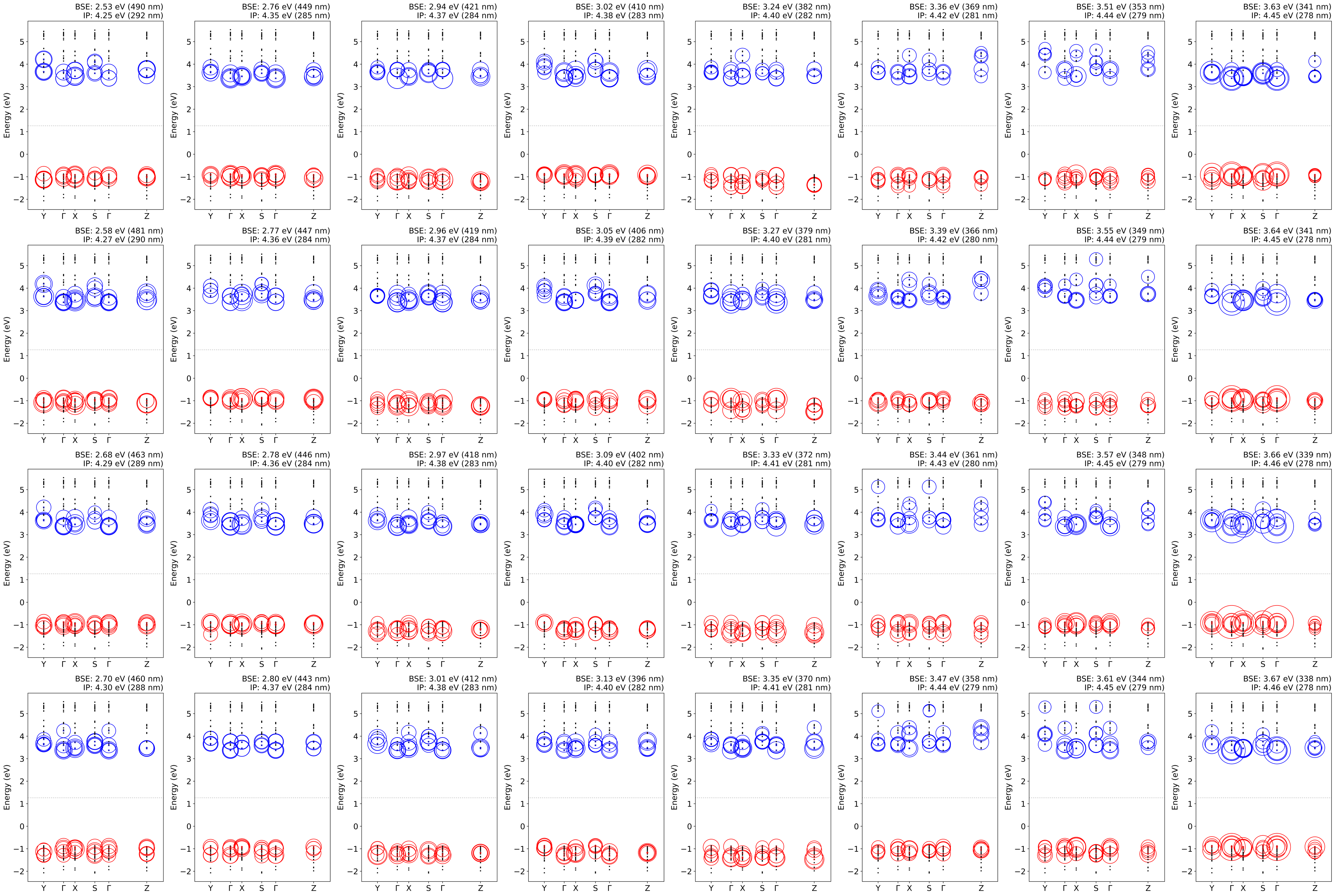}
\caption{Illustration of the $e$--$h$ coupling constants corresponding to the 32 excitation energy of PHI-3D structure optimized PBE (GGA) functionals.}
\end{figure}
\newpage

\begin{figure}
\centering
\includegraphics[width=0.9\textwidth]{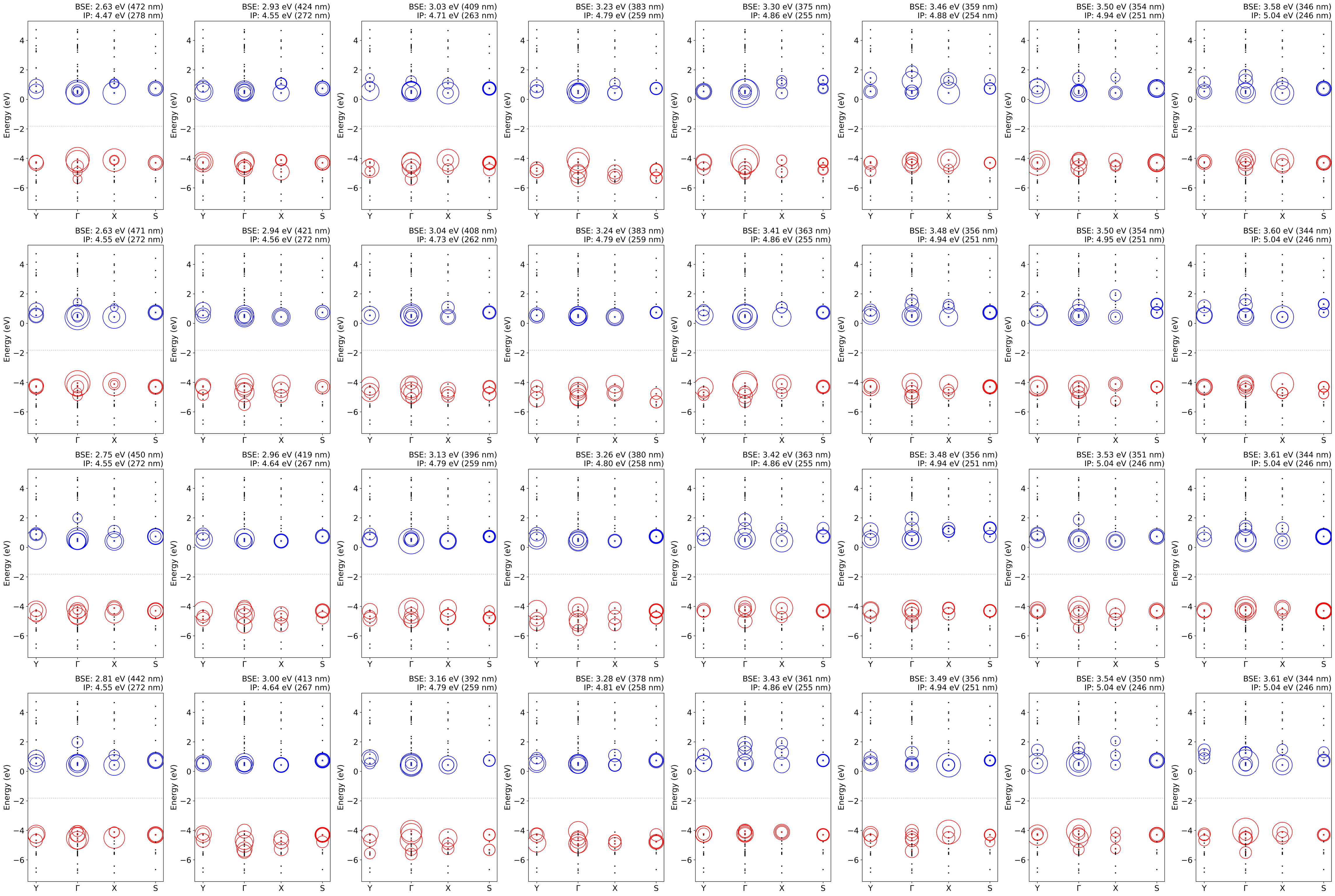}
\caption{Illustration of the $e$--$h$ coupling constants corresponding to the 32 excitation energy of c-gCN-2D structure optimized PBE (GGA) functionals.}
\end{figure}
\newpage
\begin{figure}
\centering
\includegraphics[width=0.9\textwidth]{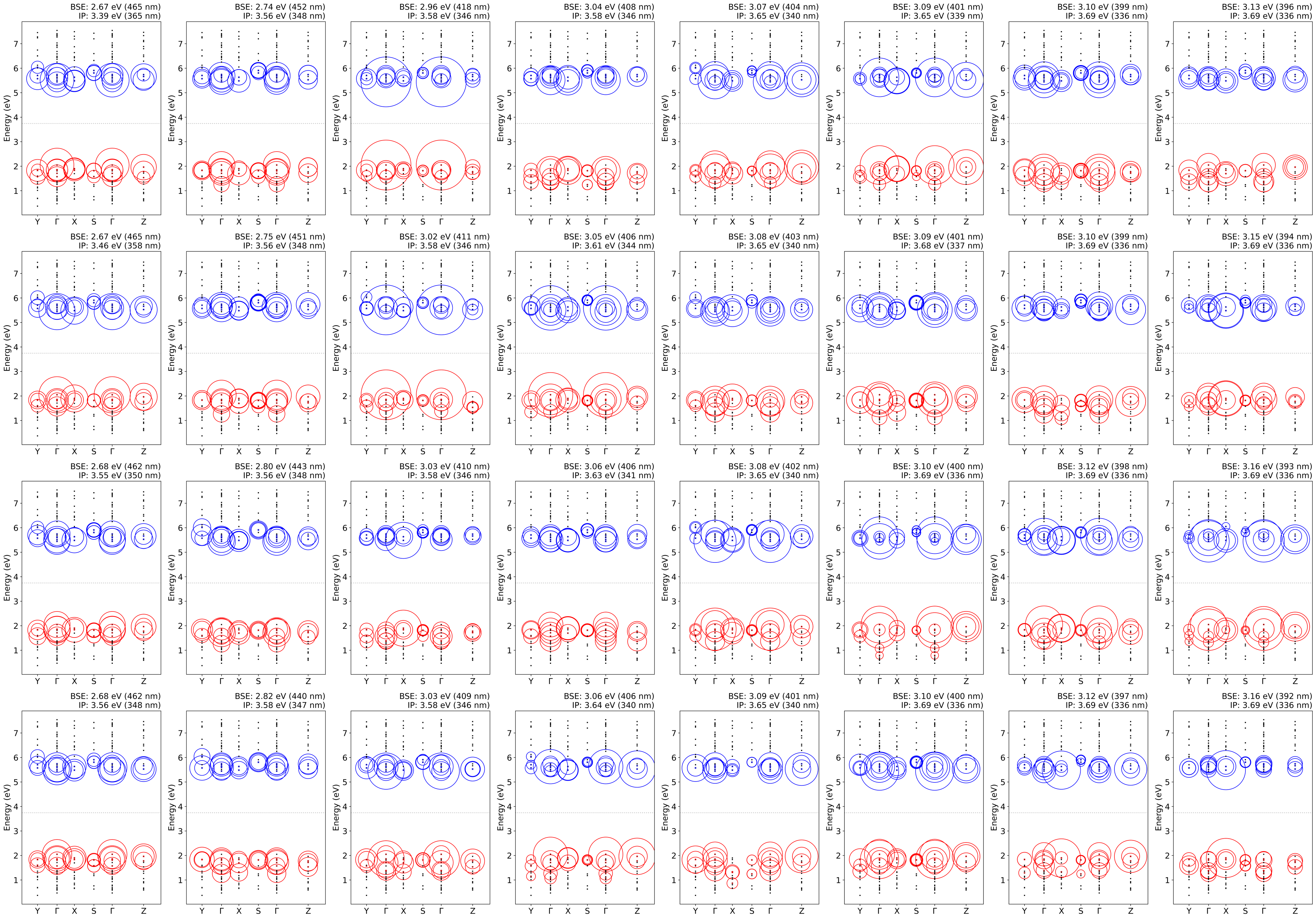}
\caption{Illustration of the $e$--$h$ coupling constants corresponding to the 32 excitation energy of c-gCN-3D structure optimized PBE (GGA) functionals.}
\end{figure}

\begin{figure}
\centering
\includegraphics[width=0.9\textwidth]{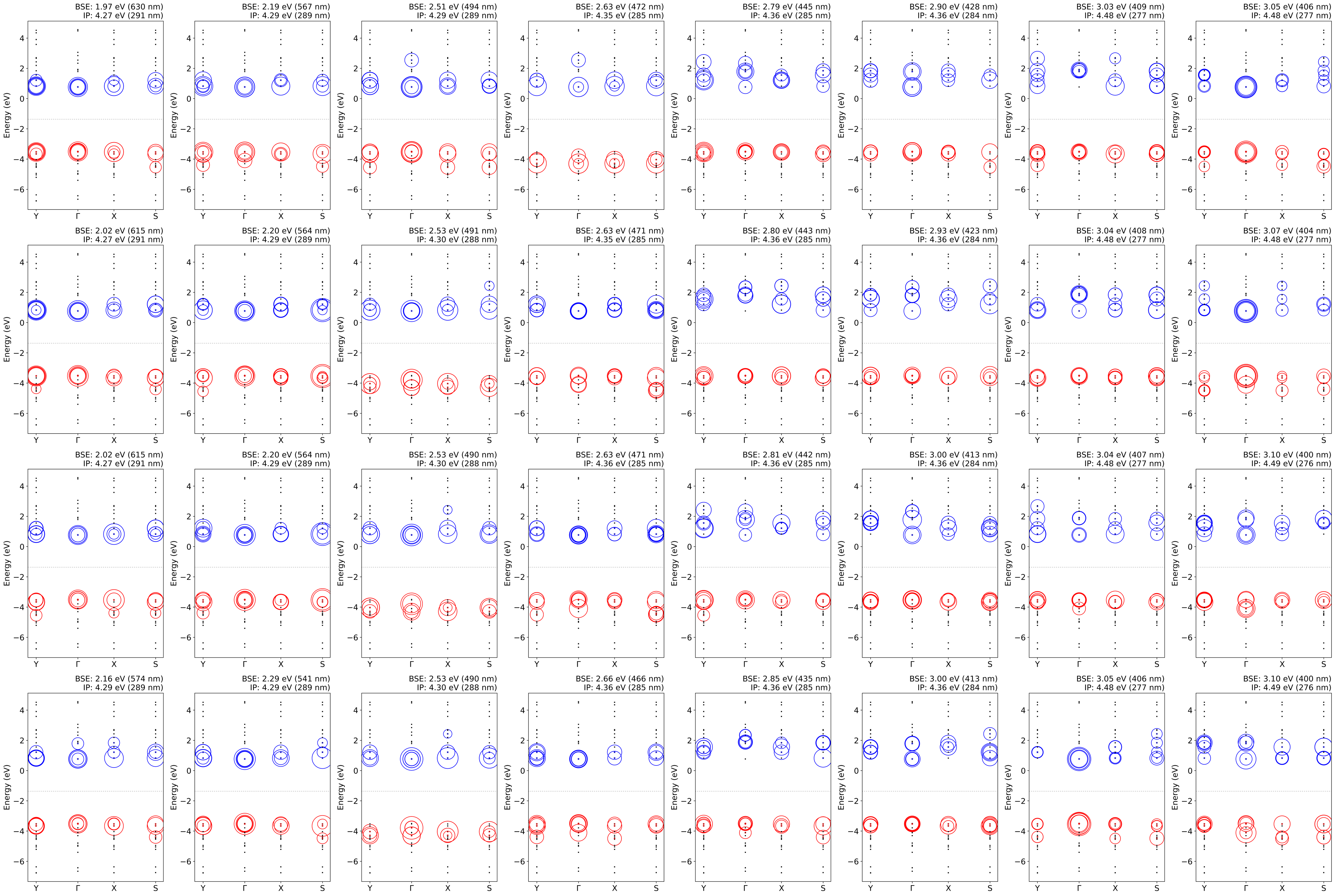}
\caption{Illustration of the $e$--$h$ coupling constants corresponding to the 32 excitation energy of p-gCN-2D structure optimized PBE (GGA) functionals.}
\end{figure}

\begin{figure}
\centering
\includegraphics[width=0.9\textwidth]{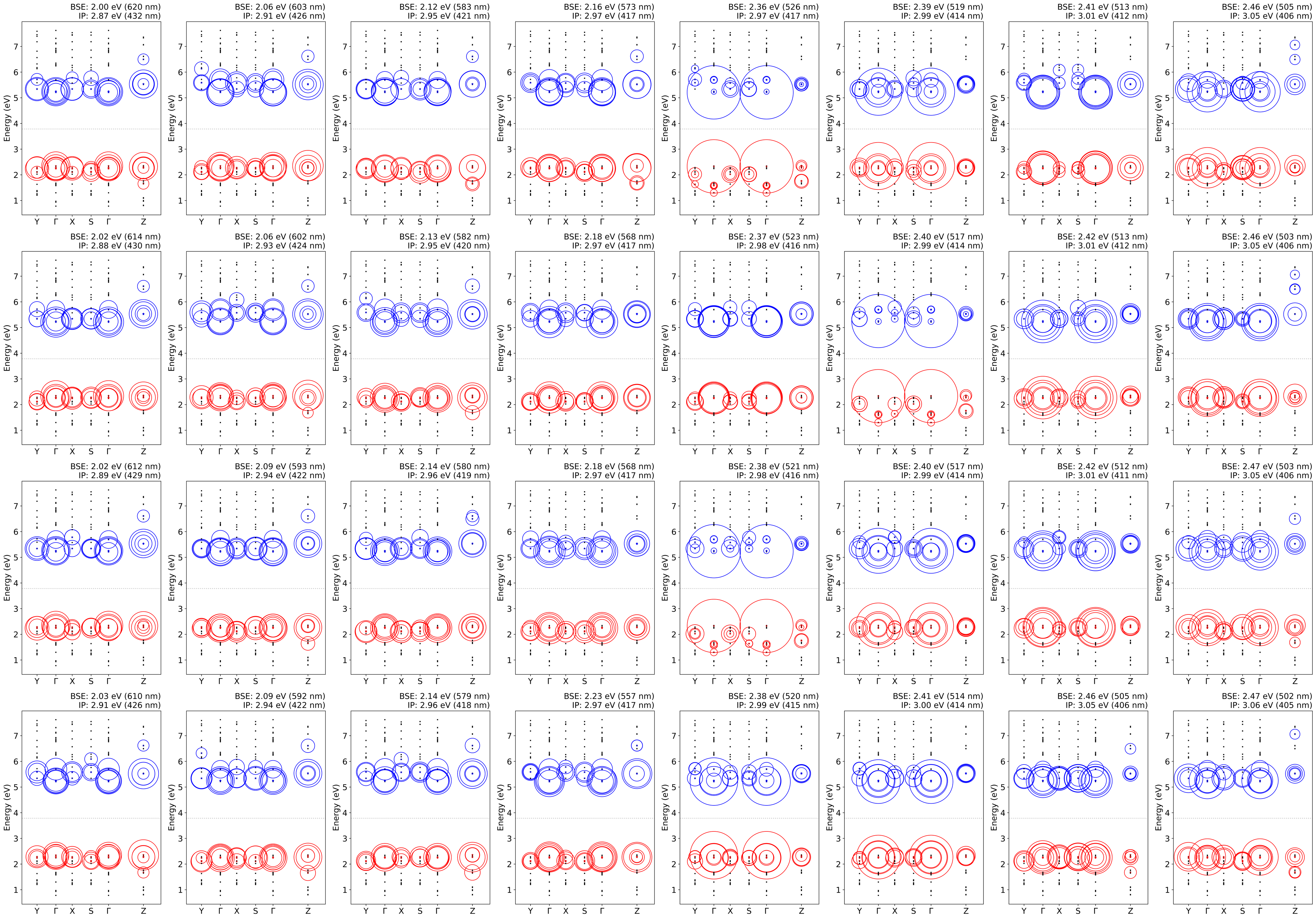}
\caption{Illustration of the $e$--$h$ coupling constants corresponding to the 32 excitation energy of p-gCN-3D structure optimized PBE (GGA) functionals.}
\end{figure}

\end{document}